\documentclass[11pt,a4paper]{article}
\usepackage{amsmath,amssymb,amsfonts,amsthm}
\usepackage[hidelinks,breaklinks=true]{hyperref}
\usepackage{authblk}
\usepackage[english]{babel}
\usepackage[utf8]{inputenc}
\usepackage{xcolor}
\usepackage{caption}
\usepackage{sectsty}
\usepackage{subcaption}
\usepackage{graphicx}
\usepackage{dcolumn}
\usepackage{bm}

\subsectionfont{\normalfont\scshape}
\subsubsectionfont{\normalfont\itshape}

\newcommand{\R}{\mathbb{R}}
\newcommand{\Ax}{\mathcal{A}}

\newcommand{\N}{\mathcal{N}}
\newcommand{\Ho}{\mathcal{H}}
\newcommand{\M}{\mathcal{M}}
\newcommand{\dom}{\mathcal{D}}
\newcommand{\bi}{\begin{itemize}}
\newcommand{\ei}{\end{itemize}}
\newcommand{\be}{\begin{equation}}
\newcommand{\ee}{\end{equation}}
\newcommand{\rhomax}{\rho_{\rm MAX}}
\newcommand{\bsigma}{{\bar\sigma}}
\newcommand{\bomega}{{\bar\omega}}

\newcommand{\Sf}{\mathcal{S}}
\newcommand{\qsf}{\mathfrak{q}}

\newtheorem{definition}{Definition}

\newtheorem{problem*}{Problem}
\newtheorem{proposition}{Proposition}

\usepackage{caption}
\usepackage{comment}

\begin{document}
\title{\bf\Large Periodic Analogs of Multiple Black Holes Solutions}

\author[1]{Omar E. Ortiz\footnote{omar.ortiz@unc.edu.ar}}
\affil{FAMAF, Universidad Nacional de Córdoba, Argentina}
\author[2]{Javier Peraza\footnote{jperaza@pitp.ca}}
\affil{Perimeter Institute for Theoretical Physics, 31 Caroline St. N., Waterloo ON, Canada, N2L 2Y5}

\date{}

\renewcommand\Affilfont{\itshape\small}
\maketitle

\begin{abstract}
In this article, we extend the numerical studies developed in \cite{Peraza:2023xic} to construct periodic stationary axisymmetric solutions containing multiple horizons in each fundamental domain. As a direct application, we consider periodic stationary axisymmetric solutions with two identical equidistant counter-rotating horizons. These solutions can be parametrized by the period $L$, the mass $M$, and the absolute value of the angular momentum $|J| >0$. We provide strong numerical evidence for the existence of such configurations, without any restriction in terms of the distance between horizons. This is in sharp contrast with the non-zero total angular momentum case, as it was recently established in \cite{Peraza:2024uto} that static single-horizon periodic solutions cannot be put into rotation if $L < 4M$. It is shown that these solutions do not have any struts on the axis, and it is also explicitly shown that, by taking non-equidistant horizons, struts develop between the black holes. Other global properties of the solutions are also presented.
\end{abstract}

\section{Introduction}

The classification of equilibrium (stationary) configurations for multiple black holes is an important problem in General Relativity. For the standard topology and asymptotically flat case, these configurations arise as the usual heuristic picture in the late-time stage of any gravitational collapse of an isolated system, where a stationary regime is assumed, resulting in one or several rotating black holes. A very closely related question to this problem is whether the family of Kerr solutions is unique among asymptotically flat stationary black hole solutions with finitely many disconnected horizons.

Particular configurations of stationary black holes have been studied in recent decades. The case of binary solutions with two coaxial rotating black holes, where there are two disconnected horizons, has been thoroughly studied using a wide range of techniques, from harmonic map analysis to the inverse scattering method and numerical simulations. The overall result is that equilibrium is generally not possible without breaking the regularity of the solution due to the development of struts (i.e., conical defects).

The axisymmetric case represents the main research avenue. Ernst \cite{Ernst:1967wx} and Carter \cite{Carter:1969zz} showed that the axisymmetric stationary Einstein equations reduce to a harmonic map from a two-dimensional domain in three-dimensional Euclidean space to the hyperbolic plane. Weinstein, in a series of works \cite{Wei90, Weinstein1992} showed that a solution to the harmonic map problem with prescribed singularities always exists (and is essentially unique), but these solutions generically have a strut between the horizons. These struts are conical singularities and therefore no solutions regular at the axis are possible. 

A large family of axisymmetric stationary black hole solutions can be explicitly constructed using the Inverse Scattering Method, \cite{Belinsky:1978nt, Belinsky:1979mh, Neugebauer:1979iw, Neugebauer:1980}. Although the construction provides the explicit functions starting from a set of \textit{coordinate} parameters (usually in terms of the Weyl-Papapetrou cylindrical coordinates), the regularity of the solution needs to be checked a posteriori. The \textit{physical} parameters, such as angular momentum or mass, depend non-trivially on the \textit{coordinate} parameters.

In \cite{Manko2008, Manko2011,Manko2017}, Manko \textit{et al.}, constructed explicit exact solutions with two horizons, i.e., binary stationary black hole systems, both counter-rotating and co-rotating along the axisymmetric axis, kept apart from falling onto each other by a massless strut. Neugebauer and Henning showed, in \cite{Neugebauer2009}, the non-existence of solutions with two non-degenerate horizons using the inverse scattering method by showing a contradiction with the inequality between area and angular momentum \cite{Dain:2011pi}. Little work has been done on cases with three or more disconnected horizons.

During the last decades, however, black hole solutions with non-standard topologies or non-standard asymptotic behavior have gained considerable attention, as well as the search for new solutions in higher dimensions \cite{Myers1987, Korotkin1994, Korotkin:1994cp, Elvang:2007rd, Emparan:2001wn,Khuri:2020dbw, Khuri:2021fqu}. Periodic static solutions in 3+1, where infinitely many coaxial black holes are stacked in an array along the axis (i.e., a linear \textit{pearl necklace}-like distribution), have both a non-standard topology (they are asymptotically $\mathbb{R} \times S^1 \times S^1$) and a Kasner-like asymptotic behavior. These solutions were discovered independently by Myers \cite{Myers1987} and Korotkin and Nicolai \cite{Korotkin:1994cp}. They are referred to as ``periodic Schwarzschild'' solutions since there is a single static horizon in the fundamental domain of the solution.

A natural question relevant to both the context of the classification of stationary solutions and the existence of solutions with non-standard topologies, is whether a ``periodic Kerr'' solution exists. It was first conjectured to exist in \cite{Korotkin:1994cp}. Recently, this problem has been studied using several techniques, from numerical constructions to geometric analysis. In \cite{Peraza:2023xic} a numerical study was carried out to construct solutions with periodic, coaxial, equidistant rotating black holes, in which each horizon has the same area $A$ and angular momentum $J$, and the separation between two consecutive horizons is $D$. In view of the numerical results and a heuristic argument, it was conjectured that periodic configurations appear to exist only when the distance between two consecutive black holes is larger than a certain critical value, depending on $A$ and $|J|$. Later, in \cite{Peraza:2024uto}, the authors formally showed that a static Myers/Korotkin-Nicolai (MKN) solution cannot be put into rotation if $12 \; D < \sqrt{A}$. This result establishes a curious static rigidity in the 3+1 solution, which is not present in known solutions.

Both results, non-existence in some regimes and numerical existence (i.e., explicit construction of some numerical examples) in other regimes, lead naturally to the study of the solution space. In \cite{Korotkin:2025pfm}, the solution space was studied in detail by means of the Inverse Scattering Method. Given a set of parameters, the numerical solution was constructed, and the corresponding Lewis-type asymptotic was obtained. If the Kasner exponent is greater than or equal to 1, the solution is not regular (a singularity emerges at a finite distance from the axis). Otherwise, the solution is regular everywhere. 

In this work, we take a step forward in the analysis of multi-horizon set-ups in periodic spacetimes. By using two black holes with opposite spins but otherwise identical, we obtain a vanishing total angular momentum. This implies that the \textit{asymptotic region} is Kasner-like at leading order, in contrast to the non-vanishing total angular momentum case, where the asymptotic region is Lewis to leading order (see \cite{Peraza:2023xic}). The main difference is that, a priori, we do not have any obstruction to existence in the asymptotic region. Only local properties of the black hole can impose a breakdown of the solutions.

We will show  that the lower bound for the area, $12 \; D < \sqrt{A}$, is not present in the periodic counter-rotating double Kerr solution and provide a detailed analysis of the solutions. We observe that in the limit $D \rightarrow 0$, the solutions are numerically unstable, indicating a possible physical obstruction. In particular, the angular velocity of each horizon seems to grow unbounded as $D \rightarrow 0$.

The paper is organized as follows. In \autoref{sec_Background} we present the problem and give the main definitions. In \autoref{sec_num} we present the setup and boundary conditions necessary for the numerical approach to the problem, as well as the construction of the initial seed. In \autoref{counterrotsection} we show the numerical results for the counter-rotating binary periodic solutions. Finally, in \autoref{sec_conlusions} we discuss the results and mention some future research lines.

\section{Background and definitions}\label{sec_Background}

A general axially symmetric stationary solution to the Einstein Equation is,
given by its line element (see for example \cite{Wald1984}),
\begin{equation} \label{asm}
ds^2 = -V dt^2 + 2 W dt d\phi + \eta d\phi^2 + e^{2u} (d\rho^2 + dz^2)
\end{equation}
where $(t, \rho ,z , \phi)$ are the Weyl-Lewis-Papapetrou coordinates. The
coordinate ranges are
\begin{equation} 
\rho \in [0 , + \infty), \quad z \in \R , \quad t \in \R , \quad \phi \in S^1,
\end{equation}
and $V(\rho ,z),W(\rho ,z),\eta(\rho ,z),u(\rho ,z)$ are functions that depend only on $(\rho ,z)$. These functions are more suitably written in terms of $\omega(\rho,z)$ and $\sigma(\rho,z)$, via the algebraic relations
\begin{align}
W &= \eta \Omega, \label{W}\\
V &= \frac{\rho^2 - W^2}{\eta} ,\\
e^{2u} &= \frac{e^{2 \gamma}}{\eta}, \\
\eta &= \rho^2 e^\sigma \label{eta},
\end{align} 
where $\Omega$ and $\gamma$ are given by the following quadratures 
\begin{align}
\Omega_z &= \rho \frac{\omega_{\rho}}{\eta^2}, & \Omega_\rho &= -\rho
\frac{\omega_z}{\eta^2}, \label{quadOmega} \\
\gamma_z &= \frac{\rho}{2 \eta^2} (\eta_\rho \eta_z + \omega_\rho \omega_z), &
\quad \gamma_\rho &= \frac{\rho}{4 \eta^2} (\eta_\rho^2 - \eta_z^2 +
\omega_\rho^2 - \omega_z^2). \label{quadgamma}
\end{align}

Einstein Equation reduces to a system of partial differential equations for the
functions $\sigma$ and $\omega$ (see e.g. \cite{stephani_kramer_maccallum_hoenselaers_herlt_2003})
\begin{align}
\Delta \sigma &= - \frac{e^{-2 \sigma} \|\partial \omega \|^2}{\rho^4}
\label{REE1} \\
\Delta \omega &= 4 \frac{\partial_i \omega  \partial^i \rho }{\rho} + 2
\partial_i \omega \partial^i \sigma \label{REE2}
\end{align}
where the inner product is the one of the flat metric $d\rho^2 + dz^2$, and the
Laplacian is the $\R^3$ Laplacian in cylindrical coordinates $(\rho,\phi,z)$
under axial symmetry:
\begin{equation} 
\Delta = \partial^2_\rho + \frac{1}{\rho} \partial_\rho + \partial^2_z
\end{equation}

To find a solution to the Einstein Equation, one first solves equations \eqref{REE1} and \eqref{REE2}, then the quadratures \eqref{quadOmega} and \eqref{quadgamma}, and finally the algebraic equations \eqref{W}-\eqref{eta}.

We are interested in regular solutions everywhere. By definition, the metric \eqref{asm} is
regular at the axis if
\begin{equation}
\lim_{\rho_0 \rightarrow 0^+} \frac{\sqrt{\eta}}{\int_0^{\rho_0} e^u d \rho } =1.
\end{equation}
This condition can be handled more easily in terms of a new function $q$,
\begin{equation}\label{qfunction}
q = u - \sigma/2,
\end{equation}
which satisfies the quadrature equations,
\be \label{quad_q}
\partial_\rho q = \frac{\rho}{4}((\partial_\rho \sigma )^2 - (\partial_z \sigma )^2) + \frac{\rho}{4 \eta^2} ( (\partial_\rho \omega )^2 - (\partial_z \omega)^2) , \quad  \partial_z q = \frac{\rho}{2} \left( \partial_z\sigma \partial_\rho \sigma + \frac{1}{\eta^2} \partial_z \omega \partial_\rho \omega \right).
\ee
Substituting this function in the above limit, we have that regularity along the axis is
achieved if 
\begin{equation}
q(\rho = 0) = 0 .
\end{equation}

\subsection{Periodic stationary and axisymmetric data}

Specific boundary conditions for $\eta$ and $\omega$ are prescribed to describe black hole solutions, as shown in \cite{Wei90}. Indeed, consider a stationary and axisymmetric black hole solution given by the metric \eqref{asm}. Let $H = \N \cap \Ho$, i.e., $\partial \N = H$. Each connected component of the horizon is denoted as $H_i$, with $i =1,...,N$. Each connected component of the axis $\Ax$ is denoted as $\Ax_i$, so that $\Ax_{i-1} \cap H_i = S_i$ is the ``south pole'' of $H_i$ and $A_i \cap H_i = N_i$ is the ``north pole'' of $H_i$. Then, 
\be 
\partial_\rho \eta \mid_{\Ax_i} (0,z) = 0, \quad \omega \mid_{\Ax_i} = c_i \in \R,
\ee
and 
\be 
\eta \mid_{H_i \setminus \{S_i,N_i\}} (0,z) > 0, \quad \partial_\rho \omega \mid_{H_i} = 0,
\ee
where the constants $c_i$ are such that the angular momentum at each $H_i$ is given by 
\be 
J_i = \frac{1}{8} \left( c_{i} - c_{i-1} \right)
\ee 

On the surface $\Sf = \R_{0,\rho}^+ \times S^1_z$, given the parameters $(\{m_i\}_{i=1}^N, \{z_i\}_{i=1}^{N})$, with $m_i> 0$, a \textit{horizon} is the union of the sets,
\be 
H_i := \{ \rho = 0  , -m_i \leq  z-z_i \leq  m_i \},
\ee
and we call it an \textit{admissible horizon} if 
\be 
H_i \cap H_j = \emptyset \quad \forall i \neq j.
\ee
The axis $\Ax$ is the following set
\be 
\Ax = \bigcup \Ax_i, \quad \Ax_i  = \{\rho = 0,  z_{i} + m_i \leq z \leq z_{i+1} - m_{i+1}\}.
\ee
The poles of each connected component are $\{ S_i := z_i - m_i , N_i := z_i +m_i\} (=  H \cap \Ax)$.
 
The parameter $L$ will denote the coordinate length of the factor $S^1_z$, i.e.,
\be 
L = \int_{S^1_z} dz.
\ee

Taking into account the previous definitions, we have the following definition for the periodic stationary data. 

\begin{definition} \label{definition_stat_data}
We call $(\Sf, \qsf , \sigma , \omega)$ a \textit{periodic stationary and axisymmetric black hole data} with parameters $(\{m_i\}_{i=1}^N, \{z_i\}_{i=1}^{N} ,L,  \{A_i\}_{i=1}^N, \{J_i\}_{i=1}^N)$, provided the following conditions hold:
\begin{itemize}
\item $H = \bigcup_i \{ \rho = 0  , -m_i \leq  z-z_i \leq  m_i \}$ is an admissible horizon.

\item The boundary conditions at $H\setminus (H \cap \Ax)$ are
\begin{equation}
\rho \partial_\rho \sigma \rightarrow 2, \quad \partial_\rho \omega = 0,
\end{equation}

\item Regularity at the axis and angular momentum of each horizon,
\begin{equation}
\text{at } \Ax: \quad \partial_\rho \sigma =0, \quad \omega = c_i,
\end{equation}
where $i$ labels each connected component of $\Ax$ and $\frac{1}{8}\left( c_{i} - c_{i-1} \right)= J_i$.

\item The metric $\qsf$ is given by 
\be 
\qsf = e^{2\gamma}(d\rho^2 + dz^2),
\ee
with $\gamma= q+ \ln\rho + \frac{1}{2}\sigma$, and $q$ satisfying \eqref{quad_q} and boundary condition $q\mid_{A_1} = 0$.
\item Area at each connected component of the horizon,
\be 
\text{Area}(H_i) := 2\pi \int_{H_i} e^{\gamma} dz = A_i.
\ee
\end{itemize}
\end{definition}

From the periodic stationary and axisymmetric black hole data, we can construct a periodic stationary and axisymmetric black hole solution, as follows,
\begin{itemize}
\item Consider the manifold
\be 
\N = \Ax \cup (S^1 \times (\Sf\setminus \Ax) ),
\ee
with the metric $h = \eta d\rho^2 + \frac{1}{\eta} \qsf$, with $\eta = \rho^2 e^{\sigma}$. 
\item Let $\Omega : \Sf \rightarrow \R$ be a solution to \eqref{quadOmega}. We define the \textit{angular velocity} of the horizon as $\Omega \mid_{\Ho_i}$, and denote is as $\Omega_{\Ho_i}$,
\begin{equation}
\Omega_{\Ho_i}: = \Omega \mid_{\Ho_i}.
\end{equation}
\item Let $\M = \N \times \R_t$ and $g$ as in \eqref{asm}.

\item Then, by the definition of the periodic stationary and axisymmetric black hole data,
\be 
\partial \N = H,
\ee
and $\chi = \partial_t + \Omega_H \partial_\phi$ is a null generator of the horizon. The temperature at each horizon will be denoted by $\kappa_i$.
\end{itemize}

We can construct a periodic stationary and axisymmetric black hole solution from the 2-dimensional data by this procedure. Observe that the prescribed singular behavior of $\sigma$ at the horizons $H_i$ arises by imposing that the lapse function $N$ for the metric \eqref{asm} vanishes on the horizon, 
\be 
\left\lbrace \begin{array}{ccc}
0 = N\mid_{\partial \N} &=& e^{-\sigma/2} \mid_{\partial \N} \\
\eta\mid_{\partial \N ^\circ} &>& 0
\end{array} \right. \quad \Rightarrow \quad  \rho \partial_\rho \sigma \mid_{\partial \N}  \rightarrow 2.
\ee

A necessary condition on the data to result in a regular solution at the axis is to ask for translational symmetries on the data. This will allow us to show the regularity of the metric at the axis, i.e., that $q=0$ at each connected component of the axis.

\begin{definition}
Given a periodic stationary and axisymmetric black hole data $(\Sf, \qsf , \sigma , \omega)$ with parameters $(\{m_i\}_{i=1}^N, \{z_i\}_{i=1}^{N} ,L, \{A_i\}_{i=1}^N, \{J_i\}_{i=1}^N)$, we say that a solution is $\omega$-even ($\sigma$-even) with respect to some $z_0$ if 
\be 
\omega (z-z_0) = \omega(-z+z_0) \quad (\sigma (z-z_0) = \sigma(-z+z_0)),
\ee
and $\omega$-odd if
\be 
\omega (z-z_0) = - \omega(-z+z_0).
\ee
\end{definition}

When we have an even number of black holes, a $\sigma$-even and $\omega$-odd data with respect to $z_0 = 0$ satisfy certain constraints on the parameters.
\be 
m_i = m_{N-i+1}, \quad z_i = L- z_{N-i+1}, \quad A_i = A_{N-i+1}, \quad J_i = J_{N-i+1},
\ee 
while $\sigma$-even and $\omega$-even data have the sign changed in the constraint on the $J$'s,
\be 
m_i = m_{N-i+1}, \quad z_i = L- z_{N-i+1}, \quad A_i = A_{N-i+1}, \quad J_i = -J_{N-i+1}.
\ee 

Observe that $\omega$-even data with respect to $z_0  =z_i$ for some $i$ implies $J_i=0$.

\begin{proposition}{\textbf{Absence of angle defects for certain configurations.}} \label{prop_no_struts}
Given a periodic stationary and axisymmetric black hole data $(\Sf, \qsf , \sigma , \omega)$ with parameters $(\{m_i/L\}_{i=1}^N, \{z_i/L\}_{i^1}^{N} , \{A_i\}_{i=1}^N, \{J_i\}_{i=1}^N)$ such that
\begin{itemize}
\item is $\sigma$-even with respect to each $z_i$ for $i=1,...,N$,
\item is $\omega$-odd with respect to each $z_i$ for $i=1,...,N$, or is $\omega$-even with respect to each $\frac{z_{i+1} + z_{i}}{2}$ for $i=1,...,N$ (with the definition $z_0 \equiv z_N$ and $z_{N+1} \equiv z_1$).
\end{itemize}
then $q = 0$ at the axis.

\begin{proof}

In view of the constraints each parity condition poses, we see that the solutions $\omega$-odd with respect to each $z_i$ for $i=1,...,N$ correspond to equidistant, identical, co-rotating black holes, while the solutions $\omega$-even with respect to each $\frac{z_{i+1} + z_{i}}{2}$ for $i=1,...,N$ correspond to equidistant, alternate spinning black hole with otherwise identical parameters.

Fix $i$. The quadratures for $q$ are \eqref{quad_q}, and a simple inspection of these equations shows that $q$ is constant on each axis component. We will show below that if $q(0,\frac{z_{i} + z_{i-1}}{2})=0$ then $q(0,\frac{z_{i+1} + z_{i}}{2})=0$.

The integral of the closed 1-form
\begin{equation}
\left( \frac{\rho}{4}( (\partial_\rho \sigma )^2 - (\partial_z \sigma )^2) + \frac{\rho}{4 \eta^2} ( (\partial_\rho \omega )^2 -
(\partial_z \omega)^2)\right) d\rho + \frac{\rho}{2} (\partial_z\sigma \partial_\rho \sigma + \frac{1}{\eta^2}
\partial_z \omega \partial_\rho \omega) dz 
\end{equation}
on the segment from $(1,\frac{z_{i} + z_{i-1}}{2})$ to $(1,\frac{z_{i+1} + z_{i}}{2})$ is zero by the symmetries of $\sigma$ and $\omega$. Also by these symmetries, the integral on the segments $[0,1]\times \{\frac{z_{i+1} + z_{i}}{2}\}$ and $[0,1]\times \{\frac{z_{i} + z_{i-1}}{2}\}$, oriented in the same direction are equal. Therefore, the integral on the three consecutive intervals is zero, so $q(0,\frac{z_{i} + z_{i-1}}{2}) = q(0,\frac{z_{i+1} + z_{i}}{2})$.
\end{proof}
\end{proposition}

\subsection{Kasner and Lewis Solutions as asymptotic models}

In a general multi-horizon setup, we classify the asymptotic behavior of the solution in terms of the total angular momentum: either $\sum_{i=1}^N J_i = 0$ or $\sum_{i=1}^N J_i \neq 0$. 

On the one hand, the vanishing of the total momentum implies that the solution should approach a non-rotating static solution in the asymptotic region. The asymptotic model for this situation is the family of Kasner solutions, usually written as:

\begin{equation}
g_K = -t^{2p_0} dt^2 + t^{2p_1}d\rho^2 + t^{2p_2}dz^2 + t^{p_3}d\phi^2
\end{equation}
where the constants $p_i$ satisfy 
\begin{equation}\label{eq:Kasner_param}
\sum_{i \geq 1} p_i = 1 + p_0, \quad \sum_{i \geq 1} (p_i)^2 = (1+p_0)^2.
\end{equation}
Via a Wick rotation (and adjusting signature), we can write the Kasner solutions in a different form:

\begin{equation}\label{kasnersol}
g_K = -\rho^{2p_1} dt^2 + \rho^{2p_0}d\rho^2 + \rho^{2p_2}dz^2 + \rho^{p_3}d\phi^2
\end{equation}

In our case, since we have an axisymmetric and stationary solution with harmonic coordinates in the transversal 2d-metric, see equation \eqref{asm}, we have that $p_2 = p_0$. This equation and \eqref{eq:Kasner_param} define a unique free parameter for the family of solutions, which we will call \emph{Kasner exponent}. If we denote by $\alpha := 2p_1$ the Kasner exponent, then

\begin{equation}
g_K = -\rho^{\alpha} dt^2 + \rho^{\alpha(2-\alpha)/2} \left( d\rho^2 + dz^2 \right) + \rho^{2-\alpha}d\phi^2 , \quad 0 \leq \alpha \leq 2.
\end{equation}

On the other hand, when we have a non-zero total angular momentum the asymptotic model is the family of Lewis solutions \cite{Lewis} (see also \cite{Stockum1937}), which are cylindrically symmetric (i.e. independent on $\phi$ and $z$) rotating stationary vacuum solutions. Some of the solutions extend to infinity ($\rho=+\infty$) and some do not. The possible forms for $\eta$ are the following
\begin{align}
\label{I} {\rm (I\pm)}:\ & \eta= \rho \frac{|w|}{a} \sin (\pm a \ln (\rho) + b), \ a> 0,\ b \in \R,\\
\label{II} {\rm (II\pm)}:\ & \eta= \rho |w| ( \pm \ln (\rho) + b),\ b \in \R,\\
\label{III} {\rm (III\pm)}:\ & \eta= \rho \frac{|w|}{a} \sinh (\pm a \ln (\rho) + b), \ a> 0,\ b \in \R,
\end{align}
where $a$ and $b$ are free parameters and $w$ is related to the twist potential by,
\be
\omega = wz,\quad w\neq 0.
\ee
See \cite{Peraza:2023xic} for a detailed analysis of these solutions. The metrics of the models (III+) are Kasner to leading order (except for the cross term $-2\text{s}(w)e^{-b}\rho^{1-a}dtd\phi$), where the Kasner exponent is $\alpha = 1-a$, and satisfies 
\begin{equation} \label{kasnersol}
g_{K} = -e^{-b}\rho^{\alpha} dt^2 + e^{b}\rho^{2-\alpha}d\phi^2 +
c\rho^{\alpha^{2}/2-\alpha} \left( d\rho^2 + dz^2 \right), \quad 0\leq \alpha \leq 1.
\end{equation}

It is important to note that a non-vanishing total angular momentum solution \emph{cannot} have a Kasner exponent greater than 1. This has major consequences for the analysis of the solution space, see \cite{Korotkin:2025pfm}.

\subsubsection{Smarr formula and Mass formula}

We refer to the following identity:
\be \label{eq:mass_formula}
\partial_\rho M(\rho)  = 0,
\ee 
as the \emph{Mass formula}, where 
\begin{equation}\label{mass_function}
M(\rho) = \left[ \frac{1}{4} \int_{-L/2}^{L/2} (- \rho \partial_\rho \sigma + \Omega \partial_z \omega ) dz \right]_{\rho}.
\end{equation}
This particular expression comes from evaluating the Komar mass integrand at some $\rho$. 

On a solution to the Einstein equations, \eqref{eq:mass_formula} holds and we can evaluate on the axis and in the limit $\rho \rightarrow +\infty$, obtaining the Smarr formula, presented in the usual form, is 
\begin{equation} \label{Smarr}
M = M( \rho \rightarrow +\infty)= M(\rho \rightarrow 0) =  \frac{1}{4 \pi} \kappa A + 2 \Omega_\Ho J, 
\end{equation}
where $M$ and $J$ are the Komar mass and angular momentum, respectively, $\kappa$ is the horizon temperature, $A$ is the horizon area and $\Omega_\Ho$ is the horizon angular velocity. The expression for both the horizon area and the surface gravity are,

\be
A = 2\pi \int_{-m}^m e^{\gamma} dz, \qquad  \kappa = -\frac{1}{2} |\nabla_\mu \xi_\nu |^2 \mid_{\Ho}
\ee
where $\xi_\nu$ is the null Killing vector field along the horizon.

\section{Numerical analysis with many horizons}\label{sec_num}

We discuss in this section the setup of the problem that we simulate in section \ref{counterrotsection} to analyze the existence of solutions with more than one connected component of the horizon per period. In particular, we are interested in the two counter-rotating black holes.

We want periodic stationary and axisymmetric black hole data to satisfy the hypotheses of Proposition \autoref{prop_no_struts} to avoid angle defects on the axis. It remains an open problem to prove whether there is any other data with no struts at the axis.

\begin{figure}
\centering
\includegraphics[scale=0.4]{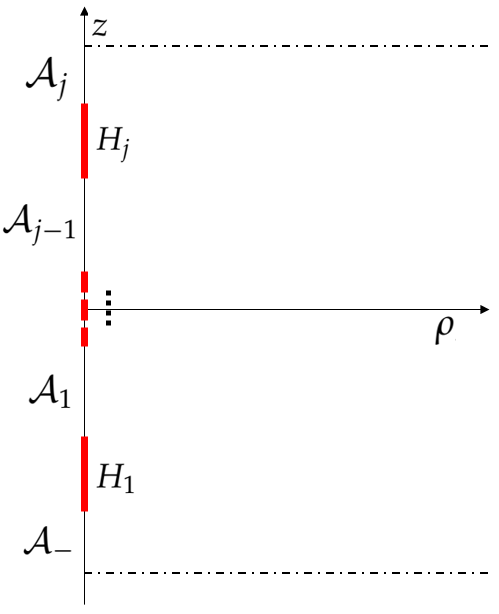}
\caption{\label{multi_horizon} Schematic representation of a multi-horizon set-up, for $j$ horizons. The data is either $z-$odd or $z-$even.}
\end{figure}

Recall that for a $z$-even or $z$-odd periodic stationary and axisymmetric black hole data on the hypotheses of Proposition \autoref{prop_no_struts}, we have that the area of each horizon is the same, they are equidistant and $|J_i|=J$ the same for all $i$. Then, once the number $j>0$ of horizons per period is fixed, the parameters $m/L,\{A_i\}_{i=1}^j , \{J_i\}_{i=1}^j$ have been set, and the parity of the solution has been decided, the boundary conditions are quite straightforward. Let 
\be 
J_T = \sum_{k=1}^j J_k, \qquad J_{\uparrow i} = \sum_{k=1}^i J_k,
\ee
be the total angular momentum of the solution with $j$ horizons per period, and the sum up to $i\leq j$, respectively. Let $\Ax_-$ be the axis component with the lowest values of $z$, $\Ax_i$ the axis component between the $i-$th and the $(i+1)-$th horizons, and $\Ax_j$ the axis component with the highest values of $z$. In \autoref{multi_horizon}, we show a diagram representing the periodic setup for the black hole data.

\vspace{0.3cm}
{\bf Boundary conditions on $\Ax$ and $\Ho$:} we distinguish between $z-$even and odd cases.

\begin{itemize}
\item If the data is $z$-even: the boundary conditions for $\omega$ are Dirichlet on the axis and Neumann on the horizon,
\be 
\omega \mid_{\Ax_-} = -4J_T, \quad \omega \mid_{\Ax_j} = -4J_T + 8J_{\uparrow j}, \quad \partial_\rho \omega\mid_{\Ho} = 0,
\ee
while for $\sigma$ are Neumann on both the axis and the horizon,
\be 
\partial_\rho \sigma \mid_{\Ax} = 0, \quad \partial_\rho (\sigma + 2 \ln \rho) \mid_{\Ho \setminus \partial \Ho} = 0,
\ee
with a Dirichlet condition at the poles, 
\be 
\lim_{(\rho,z) \rightarrow (0, z_i \pm m) }\frac{\sigma}{\sigma_i} = 1, \quad \forall 1\leq i\leq j,
\ee
where $\sigma_i$ is the reference Kerr solution with same $A$ and $J$.
\item If the data is $z$-odd: the conditions on $\sigma$ are the same, while for $\omega$ we have
\be 
\omega \mid_{\Ax_-} = 0, \quad \omega \mid_{\Ax_j} = 8J_{\uparrow j}, \quad \partial_\rho \omega\mid_{\Ho} = 0,
\ee
since $J_T = 0$ in this case.
\end{itemize}

{\bf Boundary condition at $\rhomax$:} we use analogous boundary conditions as in the one-horizon case \cite{Peraza:2023xic}. For $\omega$, we have
\be 
\omega (\rhomax , z) = \frac{8J_T}{L}z,
\ee
which, in the zero total angular momentum case, implies that $\omega(\rhomax , z) = 0$. For $\sigma$, we use the \textit{mean mass function}, $\overline{M}(\tau)$, which depends on the global properties of the intermediate states $(\sigma(\tau),\omega(\tau))$ of the flow.

Recall that by integrating by parts the last term in \eqref{mass_function}, and using \eqref{quadOmega}, we have the candidate
\begin{equation}\label{M_of_rho_at_tau}
M(\rho, \tau) = 2 \Omega(\rho, \tau) J \bigg|_{z=-L/2} - \frac{1}{4} \int_{-L/2}^{L/2}
( \rho \partial_\rho \sigma + \frac{\rho}{\eta^2} \partial_\rho \omega \omega )
dz,
\end{equation}
where $\Omega(\rho, \tau)$ is prescribed to be
\begin{equation} \label{Omega_at_tau}
\Omega(\rho, \tau) = \int_{\rhomax}^\rho \frac{\rho'}{\eta^2} \partial_z \omega d \rho', \quad \Omega (\rhomax ,z) = 0,
\end{equation}
with the functions evaluated at $z= - L/2$. That is, $\Omega$ is the integration from $\rhomax$ to $\rho$ of \eqref{quadOmega} with boundary condition $\Omega \mid_{\rhomax}(z) = 0$.

Observe that \textit{if} $(\sigma(\tau),\omega(\tau))$ converge to a solution $(\sigma, \omega)$, \textit{then} $M(\rho , \tau)$ converges to a constant function (by taking $\partial_\rho$ in \eqref{mass_function} and computing). Considering this, a natural dynamical condition arises by taking the $\rho$-\textit{average} of $M(\rho,\tau)$ over the numerical domain. At each time step $\tau$, we define  the mean renormalized Komar mass as follows, 
\begin{equation} \label{mean_M}
\overline{M} (\tau) := \overline{M(\rho,\tau)},
\end{equation}
which in particular links the values of $M$ near the axis with those far away. We use a simple, uniform average in the interior of the interval $(0,\rhomax)$\footnote{In our numerical codes, this means taking a uniform average of the values corresponding to the interior points of the discretized $\rho$ coordinate, excluding just the values at the boundary $\rho=0$ and the two grid points next to it, and the values at the outer boundary $\rho=\rhomax.$}. Then, our boundary condition for $\sigma$ at $\rhomax$ is given by,
\begin{equation} \label{asymptotic_condition_sigma}
\partial_\rho \sigma (\tau) \big|_{\rhomax} = - \frac{4 \overline{M}(\tau)}{L \rhomax}.
\end{equation} 

It is essential to notice that \eqref{asymptotic_condition_sigma} has no explicit reference to any specific Kasner exponent for the asymptotic model, nor an explicit dependence on the number of horizons at the axis. This is the main feature that allows a generalization to multi-horizon set-ups: the only assumption is $z-$independence as $\rho \rightarrow +\infty$. 

The quadrature equations \eqref{quadOmega} and \eqref{quadgamma}, for $\Omega$ and $\gamma$, respectively, are solved in the same way as in the single horizon case.

\subsection{Construction of the seed}

The construction of a general seed at $\tau = 0$ to initialize the heat flow is done analogously as in \cite{Peraza:2023xic}. Let $\sigma_{0,i}(\rho,z)$ and $\omega_{0,i}(\rho ,z)$ be the solutions to the asymptotically flat Kerr black hole with identical area $A$ and angular momentum $J_i$, located at $\Ho_i = \{|z-z_i|\leq m\}$ with $z_i = -L/2 + L/2j + (i-1)L/j$. Consider
\be \label{sum_Kerrs_multi_ho}
\sigma_0(\rho ,z) = \sum_{i=1}^j \sigma_{0,i}(\rho,z), \quad \omega_0(\rho ,z) = \sum_{i=1}^j \omega_{0,i}(\rho,z) + C_0,
\ee
where $C_0$ is such that $\omega_0$ satisfies the Dirichlet conditions defined above, given the data is either $z-$even or $z-$odd. Next, we use the same splitting as in \eqref{sig_om_desc},
\begin{align}
\sigma = & \sigma_0 + \sigma_r + \bar{\sigma},\\
\omega = & \omega_0 + \omega_r + \bar{\omega},
\end{align}
for $\sigma_r$ and $\omega_r$ defined as follows
\begin{align}
\sigma_r (\rho , z) = &\ C +  \sum_{n=1}^{\infty} \left( \sigma_0(\rho,z - nL
,J) + \sigma_0(\rho,z + nL,J) - \frac{4jM_0}{nL} \right),\\
\omega_r (\rho , z) = & \sum_{n=1}^{\infty} \left( \omega_0(\rho,z - nL ,J) + \omega_0(\rho,z + nL,J) \right),
\end{align}
where $C$ is, again, the constant such that $\sigma_r \mid_{(\rho = 0, z_i \pm m)}  =0$ for all $i = 1,..., N$. Observe that $\frac{4jM_0}{nL}$ fulfills the same role as $\frac{4M}{nL}$ in the case of a single horizon.

For $j=2$, we can, in principle, use any pair of functions $(\sigma_0, \omega_0)$ that represent a binary Kerr solution. In particular, the binary solutions presented in \cite{Cabrera-Munguia:2012gna, Cabrera-Munguia:2013jha, Cabrera-Munguia:2017dol, Manko2008,Manko2011,Manko2017} constructed via the Tomimatsu method \cite{Tomimatsu:1979rr, Tomimatsu:1981td} can be another candidate as $(\sigma_0, \omega_0)$. For $j>2$, solutions constructed with the Inverse Scattering Method can also be taken as $(\sigma_0, \omega_0)$. Heuristically, one of the advantages of working with an already binary Kerr solution would be an initially \textit{closer} expression for the asymptotic Kasner exponent already in the seed, $4M_{binary}/L$ instead of $8M_0/L$. The general feature of the metrics constructed in this fashion is that they are not regular in at least one component of the axis due to the presence of struts, and their expressions are algebraically more costly to manipulate symbolically with the computer. Therefore, we opt for the sum of Kerr solutions to approximate the binary/multi-horizon solution on each strip, \eqref{sum_Kerrs_multi_ho}.

\section{Application: two counter-rotating horizons} \label{counterrotsection}

In this section we discuss in detail the case of a periodic solution with two equidistant horizons and vanishing total angular momentum, which we call \emph{counter-rotating horizons}. Although these solutions have similar features as the ones in the co-rotating case (where all horizons have identical angular momentum) already discussed in \cite{Peraza:2023xic}, the vanishing of the total angular momentum has global consequences, in particular at the asymptotic region, where the solutions now approach a static Kasner model, instead of a Lewis model.

The input parameters (in Weyl-Papapetrou coordinates) in this set up are four: the period, $L$, the horizons lengths, $2m$, the distance between the centers of the horizons, which in this case is simply $L/2$ , and the value of the angular momentum $J$.

Let us review the construction of a Periodic Schwarzschild solution (PS) in our binary set up first. The procedure is as follows: let $u_0(\rho,z)$ be the Schwarzschild solution centered at the origin, and define the functions $u_{PS}^-$ and $u_{PS}^+$ by the series:
\begin{equation}
u_{PS}^{\pm} (\rho ,z) = u_0 (\rho,z \pm L/4) + \sum_{n=1}^{\infty} 
u_0(\rho,z - nL \pm L/4) + u_0(\rho,z + nL \pm L/4) + \frac{4M}{nL} 
\end{equation} 
The convergence and regularity of these series is immediate from the results in \cite{Korotkin:1994cp}. Therefore, both solutions are representing a periodic Schwarzschild solution, the black holes centers being at a distance $L$ from each other. Adding up both of these solutions, we obtain a chain of paired black holes, each pair consists of two periodic analogue of the Schwarzschild black hole, at a distance $L/2$, and the center of mass of each pair is a distance $L$ from each other.

Let 
\begin{equation}\label{KNMsol}
u_{PS} = u_{PS}^- + u_{PS}^+
\end{equation}
be the \textit{double periodic Schwarschild solution}. Following \cite{Korotkin1994}, it is straightforward to verify that the asymptotic behaviour of $u_{PS}$ is given by,
\begin{equation}
u_{PS} = 8M/L \ln \rho + O(1).
\end{equation}
This fall off will be important in the analysis of the asymptotic behaviour of
our numerical computations. 

Turning back the attention to our problem, we emphasize that equations
\eqref{REE1}-\eqref{REE2} are obtained from a variational principle. They
are precisely the Euler-Lagrange equations of the mass functional given by 
\begin{equation}\label{mass_functional}
\M(\sigma ,\omega)= \frac{1}{32 \pi} \int_\dom \left( |\partial \sigma |^2 +
\rho^{-4} e^{-2\sigma}|\partial \omega |^2 \right) d\mu
\end{equation}

Following \cite{Dain2009, Peraza:2023xic}, we compute the solutions to \eqref{REE1}-\eqref{REE2} via the parabolic flow
\begin{align}
\dot{\sigma} &=  \Delta \sigma + \frac{e^{-2 \sigma} \|\partial \omega
\|^2}{\rho^4}, \label{PF1} \\
\dot{\omega} &= \Delta \omega - 4 \frac{\partial_i \omega  \partial^i \rho
}{\rho} - 2 \partial_i \omega \partial^i \sigma. \label{PF2}
\end{align}
We expect that the solution of the flow will reach a stationary regime
($\dot{\sigma} = \dot{\omega} = 0$) as $t\to\infty$, thus providing a solution
to \eqref{REE1} and \eqref{REE2}.

The starting point for the parabolic flow is some arbitrary initial data $(\sigma^{(0)},\omega^{(0)}),$ at $t=0$, which we call \textit{initial seed}, that satisfies the boundary conditions for some fixed values $L$, $m$ and $J$. Then we solve numerically the flow equations \eqref{PF1} and \eqref{PF2}.

To handle the singular behaviour of the solutions at the horizons, we decompose
$\sigma$ and $\omega$ as in the previous section: we write the functions $\sigma$ and $\omega$ as a sum of known solutions to the non-periodic problem function plus a perturbation $\bar{\sigma}, \bar{\omega}$. Let $\sigma_0
(\rho,z,J)$ and $\omega_0 (\rho ,z ,J)$ be the solutions to the Kerr black hole
with momentum $J>0$ centered at the origin, and define
\begin{equation}\label{sig_om_desc}\begin{split}
\sigma =& \sigma_- + \sigma_+ + \sigma_r + \bar{\sigma} \\
\omega =& \omega_- + \omega_+ + \omega_r + \bar{\omega} 
\end{split}
\end{equation}
where
\begin{align} 
\sigma_- (\rho , z) &= \sigma_0(\rho,z - D/2, -J), \nonumber \\
\sigma_+ (\rho , z) &= \sigma_0(\rho,z + D/2, J), \nonumber \\
\omega_-(\rho ,z ) &= \omega_0(\rho,z - D/2, -J), \nonumber \\
\omega_+(\rho ,z ) &= \omega_0(\rho,z + D/2, J),  \nonumber
\end{align}
and
\begin{align}
\sigma_r (\rho , z) &= \sum_{n=1}^{N_d} \left( \sigma_0(\rho,z - nL -D/2
,-J)  + \sigma_0(\rho,z + nL - D/2 ,-J) - \frac{4M}{nL} \right)  \nonumber \\
&\quad + \sum_{n=1}^{N_d} \left( \sigma_0(\rho,z - nL +D/2 ,J) +
 \sigma_0(\rho,z + nL + D/2 ,J) - \frac{4M}{nL} \right), \nonumber \\
\omega_r (\rho , z) &= \sum_{n=1}^{N_d} \left( \omega_0(\rho,z - nL -D/2
,-J)  + \omega_0(\rho,z + nL - D/2 ,-J) \right) \nonumber \\
&\quad  + \sum_{n=1}^{N_d} \left( \omega_0(\rho,z - nL +D/2 ,J)  +
\omega_0(\rho,z + nL + D/2 ,J) \right),\nonumber
\end{align}
where $N_d \gg 1$ is some numerical cut-off parameter, and corresponds to the number of ``domains'' we stack on both the top and below the central domain. As we mentioned before, the constants $\frac{4M}{nL}$ are needed for the series to be convergent in the limit $N_d \rightarrow \infty$ (since asymptotically, each term goes as $- \frac{2M}{\sqrt{(x-nL)^2 + \rho^2}}$ and therefore we need to cancel out this divergent term). The ideal, periodic, situation occurs when $N_d \to \infty$, and the initial seed functions $\sigma^{(0)} = \sigma_- + \sigma_+ + \sigma_r$ and $\omega^{(0)} = \omega_- + \omega_+ + \omega_r$ are periodic and $C^\infty$-smooth in the variable $z$ for $\rho > 0$. 

We expect $(\bar{\sigma},\bar{\omega})$ to be regular throughout the evolution.
By inserting the decomposition \eqref{sig_om_desc} into \eqref{PF1} and \eqref{PF2}, and using the fact that each pair $(\sigma_\pm ,\omega_\pm)$ is a solution to \eqref{REE1} and \eqref{REE2}, we obtain the equations for $\bar{\sigma}$ and $\bar{\omega}$
\begin{equation}\label{bar_sigma_eq}\begin{split}
\dot{\bar{\sigma}} &= \Delta \bar{\sigma} + \Delta \sigma_r + \frac{e^{-2
\sigma_-} |\partial \omega_- |}{\rho^4} \left( e^{-2 (\bar{\sigma} + \sigma_+ +
\sigma_r )  } -1 \right) + \frac{e^{-2 \sigma_+} |\partial \omega_+ |}{\rho^4}
\left( e^{-2 (\bar{\sigma} + \sigma_- + \sigma_r)  } -1 \right) \\
&\quad + \frac{e^{-2 (\sigma_- + \sigma_+ + \sigma_r + \bar{\sigma}) }}{\rho^4}
\left( |\partial \omega_r|^2 + |\partial \bar{\omega}|^2 + 2 \left( \partial_i
\omega_r \partial^i \omega_- +   \partial_i \omega_r \partial^i \omega_+
\right.\right.\\
&\quad + \left.\left. \partial_i \omega_r \partial^i \bar{\omega} +  \partial_i \omega_- \partial^i
\omega_+ +  \partial_i \bar{\omega} \partial^i \omega_- + \partial_i
\bar{\omega} \partial^i \omega_+ \right) \right),
\end{split}
\end{equation}
\begin{equation}\label{bar_omega_eq}\begin{split}
\dot{\bar{\omega}} &=  \Delta \bar{\omega} + \Delta \omega_r - \frac{4}{\rho}
\left( \partial_\rho \omega_r + \partial_\rho \bar{\omega} \right) - 2 \left(
\partial_i \omega_- \partial^i \sigma_+ + \partial_i \omega_- \partial^i
\sigma_r + \partial_i \omega_- \partial^i \bar{\sigma} \right. \\
&\quad + \partial_i \omega_+
\partial^i \sigma_- + \partial_i \omega_+ \partial^i \sigma_r + \partial_i
\omega_+ \partial^i \bar{\sigma} + \partial_i \omega_r \partial^i \sigma_- + \partial_i \omega_r
\partial^i \sigma_+ + \partial_i \omega_r \partial^i \sigma_r \\
&\quad \left. + \partial_i
\omega_r \partial^i \bar{\sigma} + \partial_i \bar{\omega}  \partial^i
\sigma_- + \partial_i \bar{\omega} \partial^i \sigma_+ + \partial_i \bar{\omega}
\partial^i \sigma_r + \partial_i \bar{\omega} \partial^i \bar{\sigma} \right).
\end{split}
\end{equation}

\subsection{Numerical Implementation}

In this subsection we present the results of our numerical computations. To this end we show a long series with $J= 0.3$, for which we vary the relative length of the horizons with respect to the period length. The numerical technique we use to do the simulations is the same we used in \cite{Peraza:2023xic}. We repeat here the needed definitions.

The finite computational domain, $[0, \rhomax]\times[-L/2, L/2]$ where the Weyl-Papapetrou coordinates $(\rho,z)$ range, is covered by the grid
\begin{equation}\label{grid}
\left\lbrace \begin{split}
\rho_i &= \frac{1}{2}\rhomax \left( 1 - \cos \Bigl( \frac{\pi}{N_\rho}i
\Bigr) \right) \quad i = 0,...,N_\rho,\\ 
z_j &= -\frac{L}{2} + \frac{L}{N_z} \Bigl( j + \frac{1}{2} \Bigr) , \quad j =
0,...,N_z-1.
\end{split} \right.
\end{equation}
This is, a $N_\rho + 1$-point Chebyshev grid discretizes the $\rho$ coordinate,
and a $N_z$ uniform grid discretized the $z$ coordinate (the $z_j$ points are
centered with respect to the $N_z$ sub-intervals).

In most runs we use $N_\rho = 80$ and $N_z=200$, and every run in the series is
easily identified by the parameter $N_h$ which amounts the number of $z$ points
covering one of the horizons. This is, the length $2m$ of each horizon is given by
\[
2m = \frac{N_h}{N_z}L.
\]

The symmetry axis, $\{ \rho=0\}$, is included in the grid while the axis $\{ z =
0\}$ is not. The horizon with $J=-0.3$ is centered at $z=L/4$, while the horizon
with $J=0.3$ is centered at $z=-L/4.$ Also, the $z$-grid is defined in such a
way that the poles $\mathcal{H}\cap \mathcal{A}$ are at the middle of two
consecutive grid points. We use pseudo spectral and spectral collocation methods
in $\rho$ and $z$ respectively.

We refer to Section 3 of \cite{Peraza:2023xic} for many details regarding the numerical implementation of the parabolic flow and boundary conditions explained in Section 3. Moreover, we keep the same gauge fixing condition\footnote{Units for $J$ are the same as units of area, $A$.} for the area, $A = 16 \pi$.

\subsection{Parabolic flow convergence to stationary state}

We check here the convergence of the time discretization scheme (Euler forward scheme), and then the convergence to stationary state of the parabolic flow. To check the convergence of the scheme we compute three versions of a solution with typical parameters, in which we fix the space grid and change only the time step. The solutions are computed until a short time, so that the parabolic flow is still evolving significantly. We compute the solution with $N_h = 40$, or $L=9.666853$ (see \autoref{table_conv_series_J03} below) using $\delta t = 10^{-4}$, $\delta t/2$ and $\delta t/4$. 

With these three solutions we compute the quotient $Q$, which indicates the order of convergence of the method. We obtain, 
\[
Q_\bsigma = \frac{\|\bsigma^{\delta t} - \bsigma^{\delta t/2}\|}{\|\bsigma^{\delta t/2} -
\bsigma^{\delta t/4}\|} = 2.000009, \quad Q_\bomega = \frac{\|\bomega^{\delta t} -
\bomega^{\delta t/2}\|}{\|\bomega^{\delta t/2} - \bomega^{\delta t/4}\|} =
1.999958
\]
As expected, these quotients are very close to 2, since the Euler method is of
first order of convergence. 

\begin{figure}[h!]
\captionsetup{margin=1cm}
\centering
\includegraphics[width=6cm]{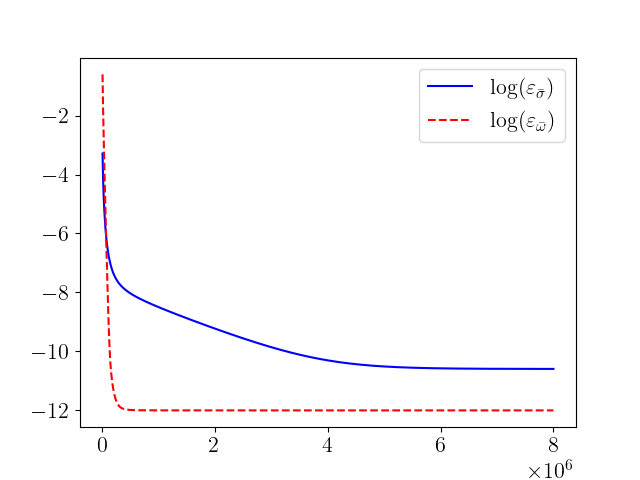}
\includegraphics[width=6cm]{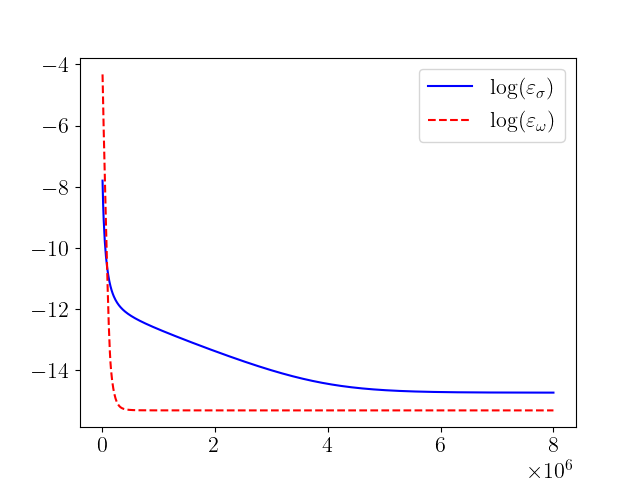}
\caption{\label{pflow_convergence_J03} Convergence of the parabolic flow to stationary state as
function of time steps for the solution with $N_h = 50$ of table
\ref{table_conv_series_J03}.}
\end{figure}

The parabolic flow evolves to stationary state in an asymptotic manner. The proximity to stationary state can be evaluated by relative errors
\begin{gather}
\varepsilon_{\bar \sigma}=\frac{\|\mbox{rhs}(\bar\sigma)\|}{\|\bar\sigma\|}, \quad
\varepsilon_{\bar\omega} = \frac{\|\mbox{rhs}(\bar\omega)\|}{\|\bar\omega\|},
\nonumber \\
\label{convergence_error} \\
\quad \varepsilon_{\sigma}=\frac{\|\mbox{rhs}(\bar\sigma)\|}{\|\sigma\|},
\quad \mbox{and} \quad \varepsilon_{\omega} =
\frac{\|\mbox{rhs}(\bar\omega)\|}{\|\omega\|}.\nonumber
\end{gather}
where $\|\cdot \|$ is the $L_2$ norm and ``rhs'' stands for ``right hand side'' of the equations \eqref{bar_sigma_eq} and \eqref{bar_omega_eq}. This is, we evaluate the relative violation of the elliptic equations for $\bar\sigma$  and $\bar\omega$. As an example, we show in \autoref{pflow_convergence_J03} the time evolution of these errors for one of the runs in the series. It is clear from these plots that the time rate at which the parabolic flow converges to the stationary state for $\bar\sigma$ and $\bar\omega$ are very different. The dynamical boundary condition given for $\bar\sigma$ makes its decay to stationary state very slow as compared to the decay for $\bar\omega$, which has a fixed, homogeneous, boundary condition. Comparison of these plots to the equivalent ones for the co-rotant case (see Figure 3 in \cite{Peraza:2023xic}) shows that in the counter-rotant case the parabolic flow reaches the stationary state faster than in the co-rotant case.

\autoref{table_conv_series_J03} shows, for each run in the series, the
errors and regularity---explained in the next section---obtained after evolving until $t=800$ where the solutions
reached the stationary state.
\begin{table}[ht!]
\captionsetup{margin=1cm}
\begin{center}
{\footnotesize
\begin{tabular}{ccccccrc}
\hline
$N_h$ & $L$ & $\varepsilon_\bsigma$ & $\varepsilon_\bomega$ & $\varepsilon_\sigma$ & $\varepsilon_\omega$ & $\Delta q~\textrm{(two)} $ & $\Delta q~\textrm{(one)}$ \\
\hline 
22 & 17.576096 & 2.69$\times 10^{-5}$ & 1.91$\times 10^{-7}$ & 1.25$\times 10^{-7}$ & 4.02$\times 10^{-9}$ & -2.45$\times 10^{-13}$ & -4.39$\times 10^{-9}$ \\
24 & 16.111421 & 1.84$\times 10^{-5}$ & 2.71$\times 10^{-7}$ & 9.41$\times 10^{-8}$ & 6.04$\times 10^{-9}$ & -1.04$\times 10^{-13}$ & -5.16$\times 10^{-9}$ \\
26 & 14.872081 & 2.14$\times 10^{-5}$ & 4.28$\times 10^{-7}$ & 1.20$\times 10^{-7}$ & 1.00$\times 10^{-8}$ & -3.44$\times 10^{-13}$ & -6.02$\times 10^{-9}$ \\
28 & 13.809790 & 2.76$\times 10^{-5}$ & 6.39$\times 10^{-7}$ & 1.70$\times 10^{-7}$ & 1.57$\times 10^{-8}$ & -2.45$\times 10^{-13}$ & -6.95$\times 10^{-9}$ \\
30 & 12.889137 & 3.26$\times 10^{-5}$ & 8.95$\times 10^{-7}$ & 2.20$\times 10^{-7}$ & 2.30$\times 10^{-8}$ &  1.34$\times 10^{-13}$ & -7.97$\times 10^{-9}$ \\
32 & 12.083566 & 3.58$\times 10^{-5}$ & 1.19$\times 10^{-6}$ & 2.64$\times 10^{-7}$ & 3.20$\times 10^{-8}$ & -3.51$\times 10^{-14}$ & -9.10$\times 10^{-9}$ \\
34 & 11.372768 & 3.73$\times 10^{-5}$ & 1.53$\times 10^{-6}$ & 3.01$\times 10^{-7}$ & 4.27$\times 10^{-8}$ & -1.25$\times 10^{-13}$ & -1.03$\times 10^{-8}$ \\
36 & 10.740948 & 3.76$\times 10^{-5}$ & 1.91$\times 10^{-6}$ & 3.30$\times 10^{-7}$ & 5.53$\times 10^{-8}$ & -1.75$\times 10^{-14}$ & -1.17$\times 10^{-8}$ \\
38 & 10.175635 & 3.69$\times 10^{-5}$ & 2.32$\times 10^{-6}$ & 3.54$\times 10^{-7}$ & 7.00$\times 10^{-8}$ & -1.33$\times 10^{-13}$ & -1.32$\times 10^{-8}$ \\
40 & 9.666853  & 3.55$\times 10^{-5}$ & 2.79$\times 10^{-6}$ & 3.71$\times 10^{-7}$ & 8.72$\times 10^{-8}$ & -2.49$\times 10^{-13}$ & -1.48$\times 10^{-8}$ \\
42 & 9.206527  & 3.37$\times 10^{-5}$ & 3.30$\times 10^{-6}$ & 3.84$\times 10^{-7}$ & 1.07$\times 10^{-7}$ &  4.48$\times 10^{-13}$ & -1.66$\times 10^{-8}$ \\
44 & 8.788048  & 3.16$\times 10^{-5}$ & 3.87$\times 10^{-6}$ & 3.93$\times 10^{-7}$ & 1.30$\times 10^{-7}$ &  8.73$\times 10^{-14}$ & -1.86$\times 10^{-8}$ \\
46 & 8.405959  & 2.94$\times 10^{-5}$ & 4.51$\times 10^{-6}$ & 3.98$\times 10^{-7}$ & 1.57$\times 10^{-7}$ &  2.33$\times 10^{-13}$ & -2.08$\times 10^{-8}$ \\
48 & 8.055711  & 2.70$\times 10^{-5}$ & 5.23$\times 10^{-6}$ & 4.01$\times 10^{-7}$ & 1.88$\times 10^{-7}$ & -6.12$\times 10^{-14}$ & -2.32$\times 10^{-8}$ \\
50 & 7.733482  & 2.47$\times 10^{-5}$ & 6.04$\times 10^{-6}$ & 4.01$\times 10^{-7}$ & 2.24$\times 10^{-7}$ &  1.24$\times 10^{-13}$ & -2.58$\times 10^{-8}$ \\
52 & 7.436041  & 2.25$\times 10^{-5}$ & 6.96$\times 10^{-6}$ & 4.00$\times 10^{-7}$ & 2.67$\times 10^{-7}$ &  2.51$\times 10^{-14}$ & -2.87$\times 10^{-8}$ \\
54 & 7.160632  & 2.04$\times 10^{-5}$ & 8.01$\times 10^{-6}$ & 3.97$\times 10^{-7}$ & 3.18$\times 10^{-7}$ &  1.20$\times 10^{-13}$ & -3.20$\times 10^{-8}$ \\
56 & 6.904895  & 1.84$\times 10^{-5}$ & 9.21$\times 10^{-6}$ & 3.93$\times 10^{-7}$ & 3.78$\times 10^{-7}$ &  6.74$\times 10^{-14}$ & -3.55$\times 10^{-8}$ \\
58 & 6.666795  & 1.65$\times 10^{-5}$ & 1.06$\times 10^{-5}$ & 3.89$\times 10^{-7}$ & 4.50$\times 10^{-7}$ & -2.38$\times 10^{-13}$ & -3.94$\times 10^{-8}$ \\
60 & 6.444569  & 1.47$\times 10^{-5}$ & 1.22$\times 10^{-5}$ & 3.83$\times 10^{-7}$ & 5.37$\times 10^{-7}$ & -2.52$\times 10^{-13}$ & -4.37$\times 10^{-8}$ \\
62 & 6.236679  & 1.31$\times 10^{-5}$ & 1.42$\times 10^{-5}$ & 3.78$\times 10^{-7}$ & 6.44$\times 10^{-7}$ &  1.54$\times 10^{-13}$ & -4.84$\times 10^{-8}$ \\
64 & 6.041783  & 1.17$\times 10^{-5}$ & 1.65$\times 10^{-5}$ & 3.74$\times 10^{-7}$ & 7.78$\times 10^{-7}$ & -1.37$\times 10^{-13}$ & -5.35$\times 10^{-8}$ \\
66 & 5.858699  & 1.05$\times 10^{-5}$ & 1.93$\times 10^{-5}$ & 3.74$\times 10^{-7}$ & 9.43$\times 10^{-7}$ & -1.09$\times 10^{-13}$ & -5.90$\times 10^{-8}$ \\
68 & 5.686384  & 9.49$\times 10^{-6}$ & 2.26$\times 10^{-5}$ & 3.79$\times 10^{-7}$ & 1.15$\times 10^{-6}$ & -4.31$\times 10^{-14}$ & -6.48$\times 10^{-8}$ \\
70 & 5.523916  & 8.64$\times 10^{-6}$ & 2.65$\times 10^{-5}$ & 3.89$\times 10^{-7}$ & 1.39$\times 10^{-6}$ & -1.72$\times 10^{-14}$ & -7.10$\times 10^{-8}$ \\
72 & 5.370474  & 7.71$\times 10^{-6}$ & 3.25$\times 10^{-5}$ & 3.95$\times 10^{-7}$ & 1.78$\times 10^{-6}$ & -1.32$\times 10^{-13}$ & -7.73$\times 10^{-8}$ \\
74 & 5.225326  & 6.59$\times 10^{-6}$ & 4.28$\times 10^{-5}$ & 3.87$\times 10^{-7}$ & 2.44$\times 10^{-6}$ &  2.21$\times 10^{-14}$ & -8.35$\times 10^{-8}$ \\ 
76 & 5.087817  & 5.58$\times 10^{-6}$ & 5.72$\times 10^{-5}$ & 3.81$\times 10^{-7}$ & 3.41$\times 10^{-6}$ & -5.97$\times 10^{-14}$ & -8.93$\times 10^{-8}$ \\
78 & 4.957360  & 5.27$\times 10^{-6}$ & 7.03$\times 10^{-5}$ & 4.26$\times 10^{-7}$ & 4.40$\times 10^{-6}$ & -3.39$\times 10^{-14}$ & -9.39$\times 10^{-8}$ \\
80 & 4.833426  & 5.82$\times 10^{-6}$ & 7.65$\times 10^{-5}$ & 5.72$\times 10^{-7}$ & 5.07$\times 10^{-6}$ & -3.04$\times 10^{-14}$ & -9.61$\times 10^{-8}$ \\
82 & 4.715538  & 7.61$\times 10^{-6}$ & 1.16$\times 10^{-4}$ & 9.57$\times 10^{-7}$ & 8.19$\times 10^{-6}$ & -1.88$\times 10^{-13}$ & -9.48$\times 10^{-8}$ \\
84 & 4.603263  & 1.40$\times 10^{-5}$ & 3.67$\times 10^{-4}$ & 2.70$\times 10^{-6}$ & 2.91$\times 10^{-5}$ &  2.44$\times 10^{-14}$ & -1.04$\times 10^{-7}$ \\
\hline
\end{tabular}
}
\caption{\label{table_conv_series_J03} Relative error after $8\times 10^6$ time steps and violation of
regularity, $\Delta q,$ for two paths (surrounding two and one horizons), for
the solutions in the series.}
\end{center}
\end{table}

\subsection{Convergence with respect to space discretization and regularity at the axis}

Not to repeat the arguments, we refer the reader to the explanation given in \cite{Peraza:2023xic} about the theory of convergence of the numerical solution to the exact solution to the differential problem. In particular, the definition of the quotient, computed with three solutions, that approaches the value $2^k$ when the method is converging with order $k$ (see equation (50) in \cite{Peraza:2023xic} and the corresponding explanation).

We now check the convergence with respect to the space discretization, by
computing on three different grids (see \autoref{table_solutions_to_check_space_discretization} below), but with the same time discretization, for the solution corresponding to $N_h=40$ in \autoref{table_conv_series_J03}.
\begin{table}[h!]
\captionsetup{margin=1cm}
\begin{center}
{\footnotesize
\begin{tabular}{cccccc}
\hline
Label & $N_\rho$ & $N_z$ & $N_h$ & $\delta t$ &  Solution\\
\hline
1 & 40 & 100 & 20 & $3.906250\times 10^{-6}$ & $\bsigma^1$, $\bomega^1$ \\
2 & 80 & 200 & 40 & $3.906250\times 10^{-6}$ & $\bsigma^2$, $\bomega^2$ \\
3 & 160 & 400 & 80 & $3.906250\times 10^{-6}$ & $\bsigma^3$, $\bomega^3$ \\
\hline
\end{tabular}
}
\caption{\label{table_solutions_to_check_space_discretization} Three
aproximations for the solution with $L=9.666853$ to check convergence with
respect to the space discretization.} 
\end{center}
\end{table}

We obtain
\begin{equation}
\frac{\|\bsigma^1 - \bsigma^2\|}{\|\bsigma^2 - \bsigma^3\|} = 1.949, 
~(\mbox{or }k=0.96), \qquad
\frac{\|\bomega^1 - \bomega^2\|}{\|\bomega^2 - \bomega^3\|} = 10.379,
~(\mbox{or }k=3.38).
\end{equation}

Finally, to check the convergence of the full numerical scheme, we compute three
approximations to the same exact solutions, by changing the space grid and the
time space, always satisfying the CFL condition as explained in equation (52) of
\cite{Peraza:2023xic}. The parameters used are shown in
\autoref{table_solutions_to_check_full_scheme}.
\begin{table}[h!]
\captionsetup{margin=1cm}
\begin{center}
{\footnotesize
\begin{tabular}{cccccc}
\hline
Label & $N_\rho$ & $N_z$ & $N_h$ & $\delta t$ &  Solution\\
\hline
1 & 40 & 100 & 20 & $1.00\times 10^{-3}$ & $\bsigma^1$, $\bomega^1$ \\
2 & 80 & 200 & 40 & $6.25\times 10^{-5}$ & $\bsigma^2$, $\bomega^2$ \\
3 & 160 & 400 & 80 & $3.906250\times 10^{-6}$ & $\bsigma^3$, $\bomega^3$ \\
\hline
\end{tabular}
}
\caption{\label{table_solutions_to_check_full_scheme} Three
approximations for the solution with $L=9.666853$.} 
\end{center}
\end{table}

We obtain
\begin{equation}
\frac{\|\bsigma^1 - \bsigma^2\|}{\|\bsigma^2 - \bsigma^3\|} = 1.940,~(\mbox{or
}k=0.96), \qquad 
\frac{\|\bomega^1 - \bomega^2\|}{\|\bomega^2 - \bomega^3\|} = 15.864,~(\mbox{or
}k=3.99)
\end{equation}
The regularity (absence of angle defect at the axis), is tested by evaluating $\Delta q$ (we refer to explanation in section 2.4.1 in \cite{Peraza:2023xic}) along two different paths: one just surrounding the two horizons, and another just surrounding a single horizon. The value of $\Delta q$ along the path surrounding the two horizons is not a very strict test because of the periodicity of the solution. The value of $\Delta q$ along a path surrounding a single horizon is a very strict test. The values of $\Delta q$ obtained for all runs in the series are shown in \autoref{table_conv_series_J03}.

It is very interesting to check how the regularity at the axis breaks down when
we compute a solution which is not symmetric. We test this by computing a
solution where the lengths of the horizons is slightly different, and compare
this solution with the two neighbouring (symmetric) solutions in
\autoref{table_conv_series_J03}. The result is shown in
\autoref{table_conv_assymetric_sol}.
\begin{table}[h!]
\captionsetup{margin=1cm}
\begin{center}
{\footnotesize
\begin{tabular}{cccccccccc}
\hline
$N_{h0}$ & $N_{h1}$ & $L$ & $H_1$ & $H_2$ & $A_-$ & $A_1$ & $A_+$ & $\Delta q~\textrm{(two)} $ & $\Delta q~\textrm{(one)}$ \\
\hline 
38 & 38 & 10.175635 & 1.9334 & 1.9334 & 2.0606 & 4.1211 & 2.0606 & -1.33$\times 10^{-13}$ & -1.32$\times 10^{-8}$ \\
39 & 38 & 9.914721  & 1.9842 & 1.9334 & 1.5263 & 3.1544 & 1.5772 & 3.24$\times 10^{-1}$ & 3.24$\times 10^0$ \\
40 & 40 & 9.666853  & 1.9334 & 1.9334 & 1.4500 & 2.9000 & 1.4500 & -2.49$\times 10^{-13}$ & -1.48$\times 10^{-8}$ \\
\hline 
\end{tabular}
}
\caption{\label{table_conv_assymetric_sol} Violation of
regularity, $\Delta q,$ for two paths (surrounding two and one horizons), for
an asymmetric solution (shown in the central line) and comparison with two
regular solutions in the series.}
\end{center}
\end{table}

\subsection{Plot of the solutions}
\begin{figure}[h!]
\captionsetup{margin=1cm}
\centering
\includegraphics[width=6cm]{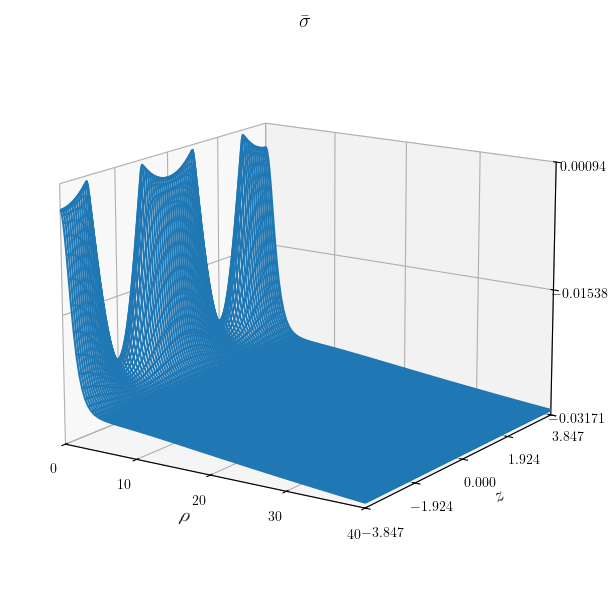}\includegraphics[width=6cm]{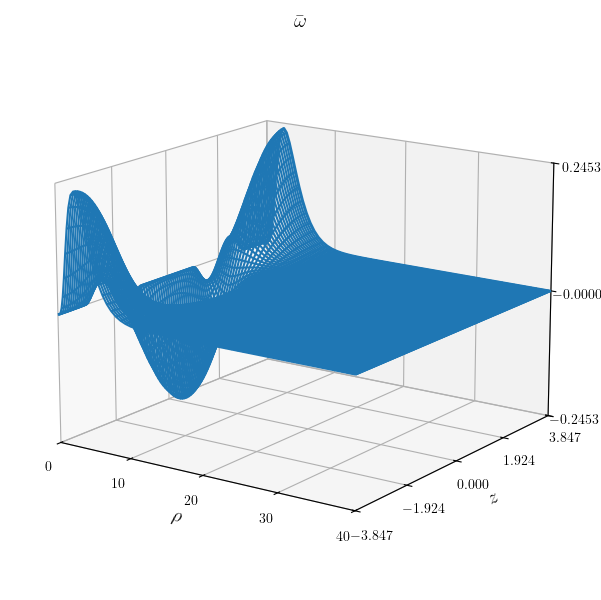}\\
\includegraphics[width=6cm]{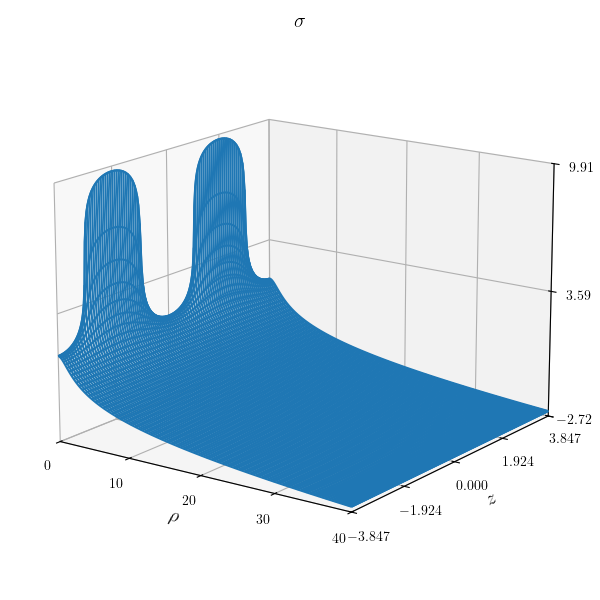}\includegraphics[width=6cm]{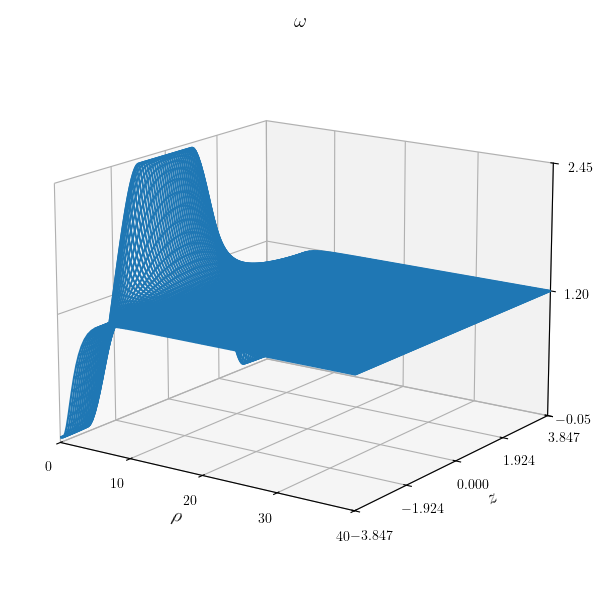}\\
\caption{\label{plot_solutions_J-03_Mh50} Plots of solutions corresponding to
$N_h=50$ of table \ref{table_conv_series_J03}. From left to right, from top to
bottom: $\bar\sigma$, $\bar\omega$, $\sigma$, $\omega$.}
\end{figure}

Here we show plots of typical solutions, see \autoref{plot_solutions_J-03_Mh50}. Although these particular graphs are for $N_h = 50$, which corresponds to $L = 7.733482$, the solutions look qualitatively similar for other values of the parameters.

\subsection{Smarr identity}
\begin{figure}[h!]
\captionsetup{margin=1cm}
\centering
\includegraphics[width=6cm]{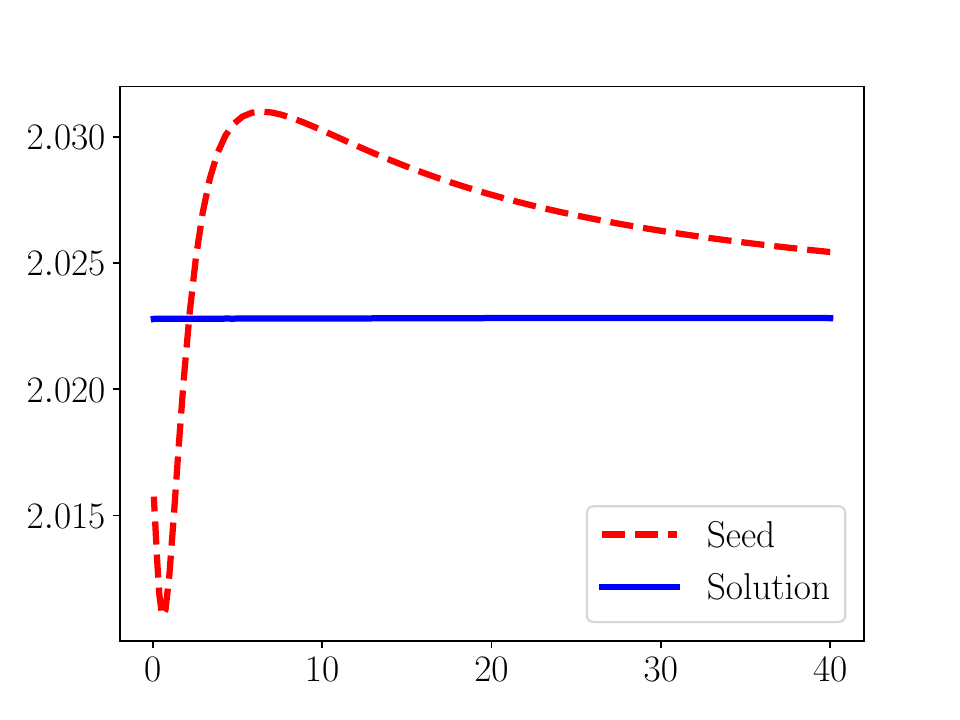}\includegraphics[width=6cm]{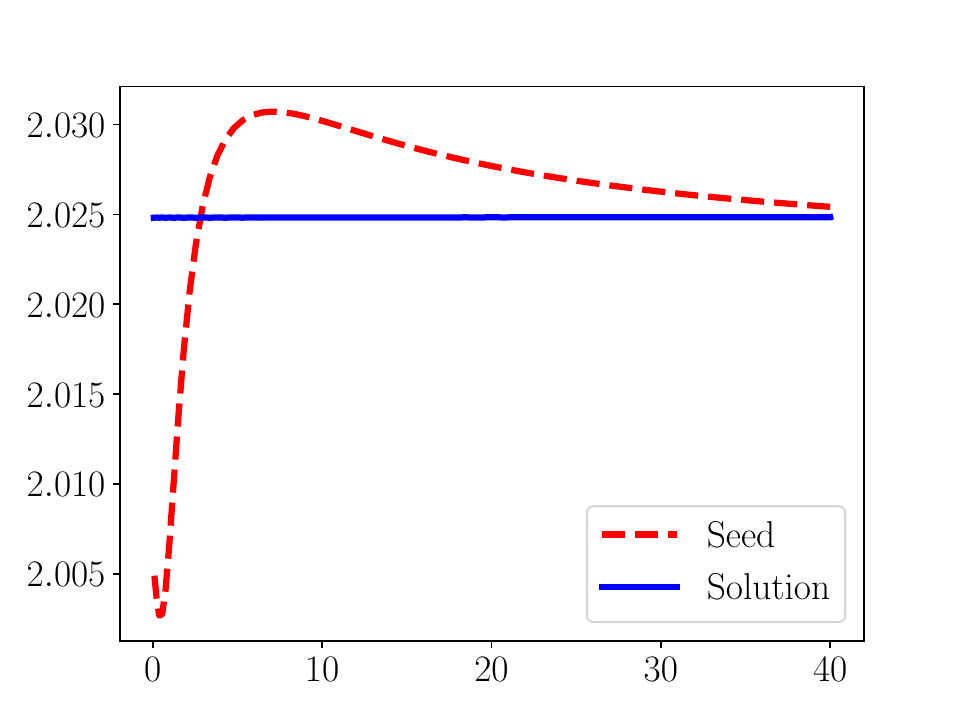}\\
\includegraphics[width=6cm]{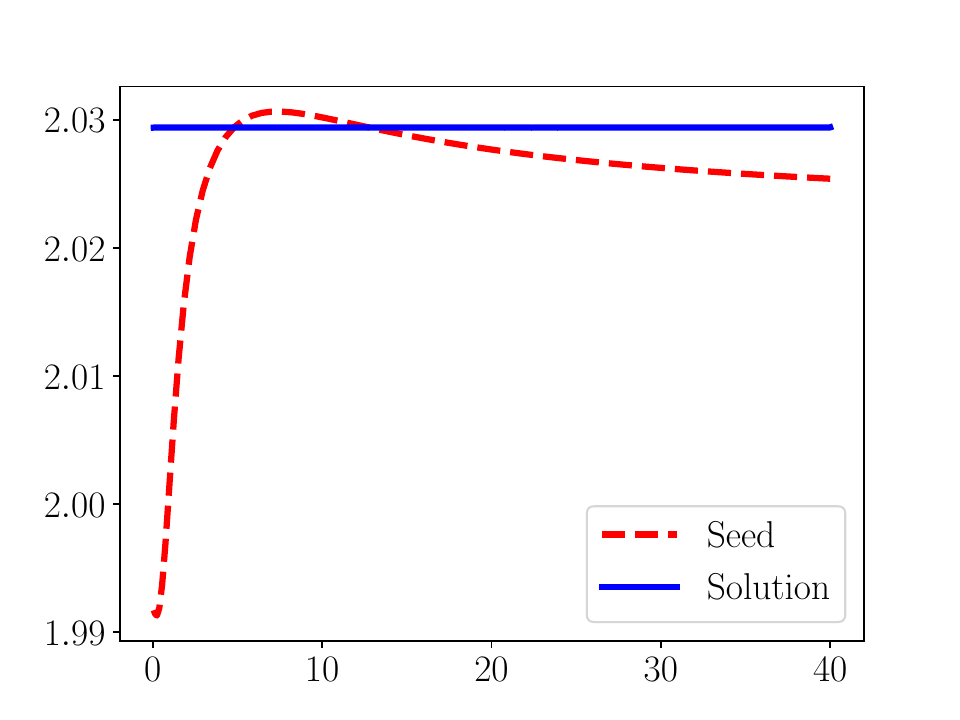}\includegraphics[width=6cm]{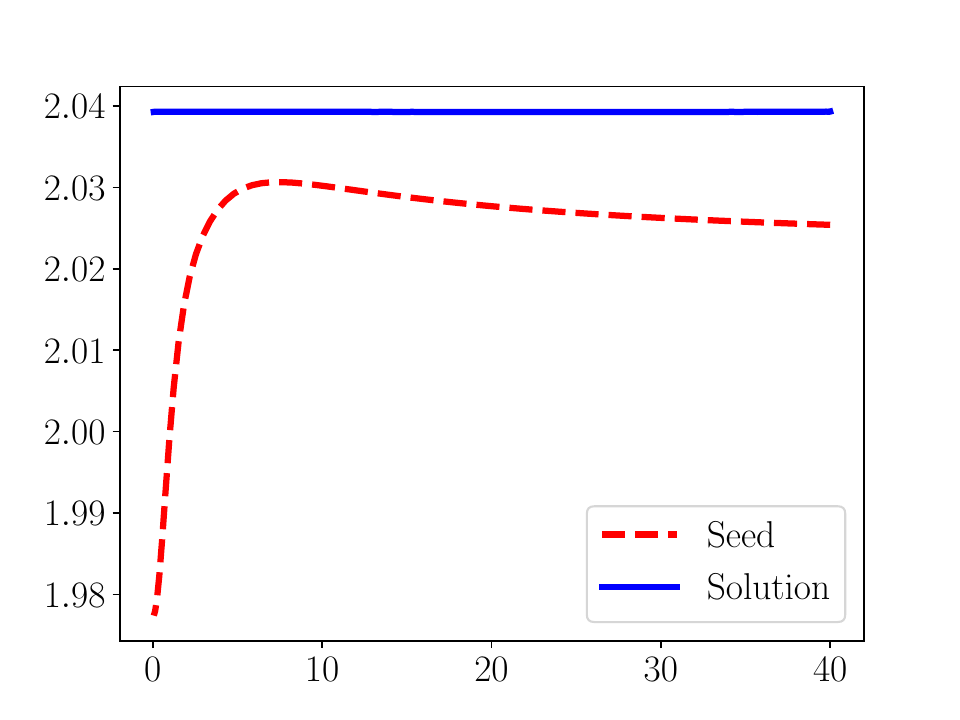}\\
\includegraphics[width=6cm]{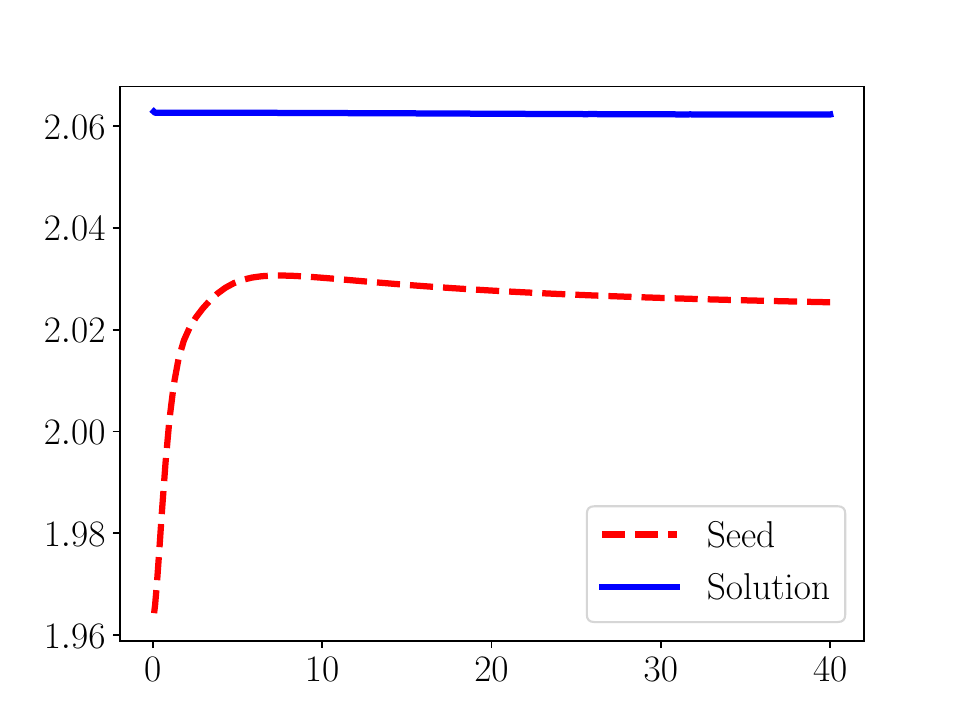}\includegraphics[width=6cm]{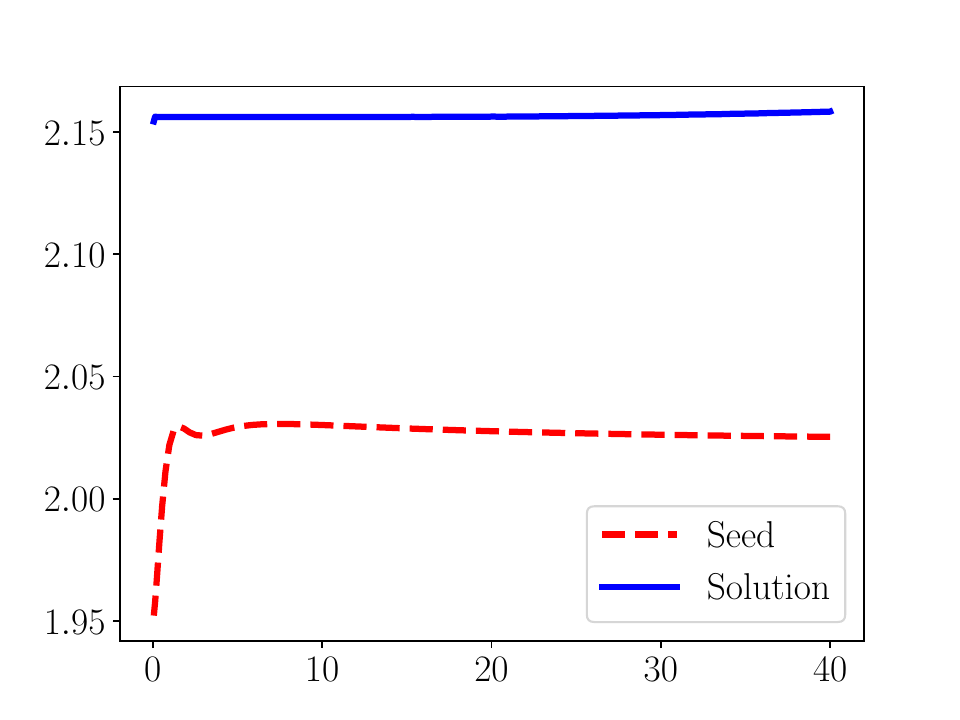}\\
\caption{Plots of $M(\rho, \tau)$ for $\tau=0$ (seed) and final $\tau$ (solution)
for six selected solutions of \autoref{table_conv_series_J03}. From left to right, from top to bottom, the solutions corresponding to $N_h = 22, 34, 46, 58,
70, 82.$}\label{figure_smarr_identity}
\end{figure}

As in \cite{Peraza:2023xic}, the validity of the Smarr identity is a very sensitive test for the solutions. In fact it is precisely this fact that allows us to find the right outer boundary condition for the $\sigma$ equation, as explained before. We show in \autoref{figure_smarr_identity} the violation of the Smarr identity by the seed (non constancy of $M(\rho,\tau=0)$), and the validity of identity for the solution (constancy of $M(\rho,\tau=0)$), for some selected solutions in \autoref{table_quantities_series_J03}.

Now that we have made all the relevant checks for the solutions, we show the
physical quantities obtained for each of them.

\subsection{Mass, Kasner parameter and angular momentum}
\begin{figure}[h!]
\captionsetup{margin=1cm}
\centering
\includegraphics[width=6cm]{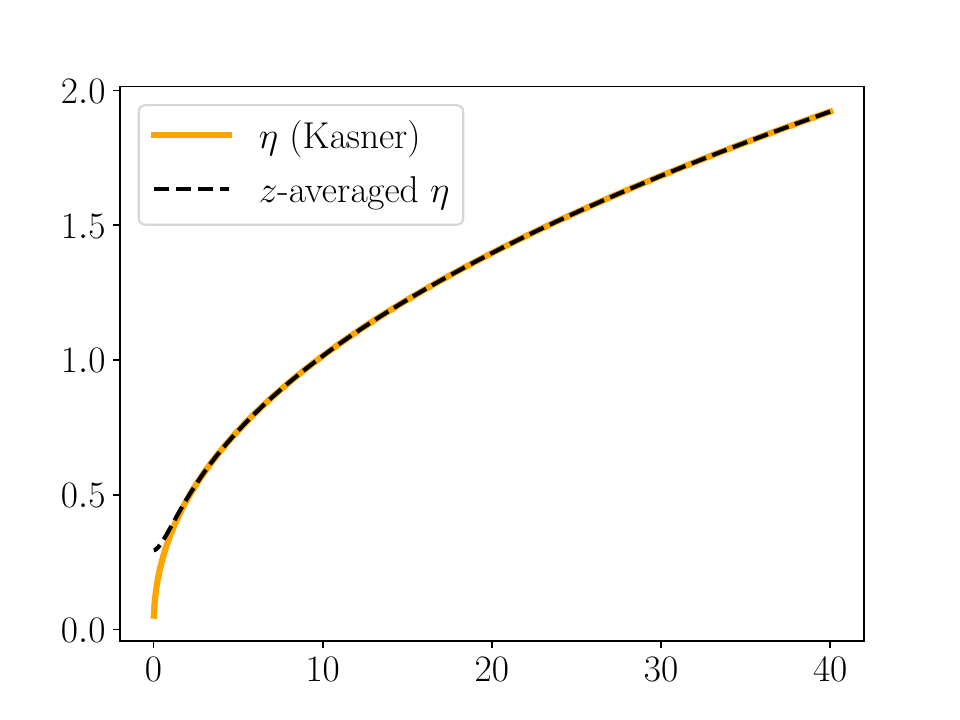}\includegraphics[width=6cm]{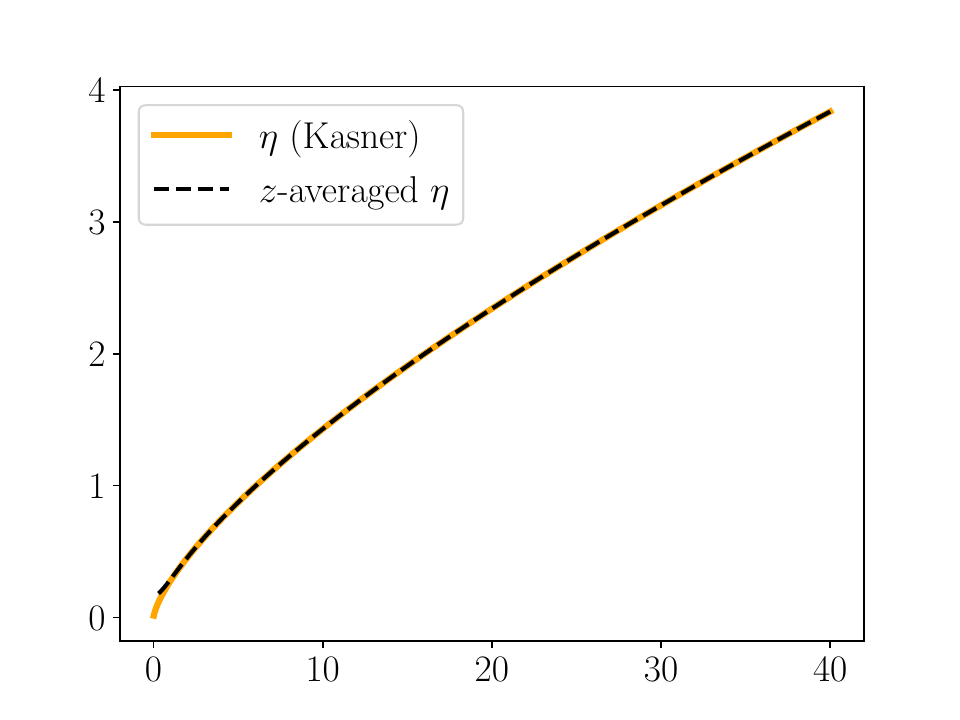}\\
\includegraphics[width=6cm]{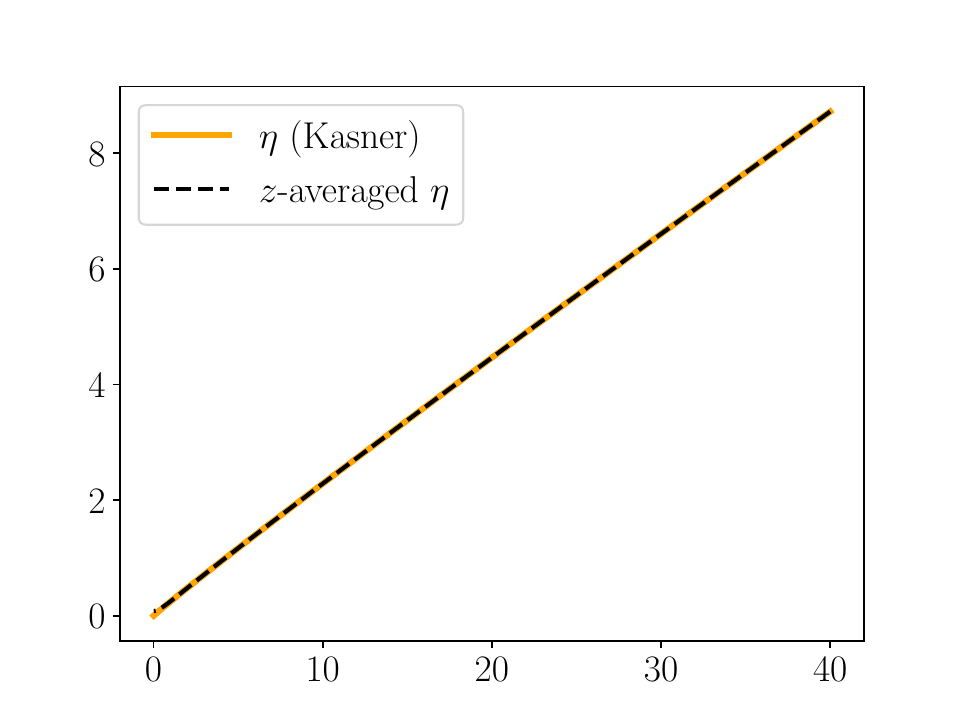}\includegraphics[width=6cm]{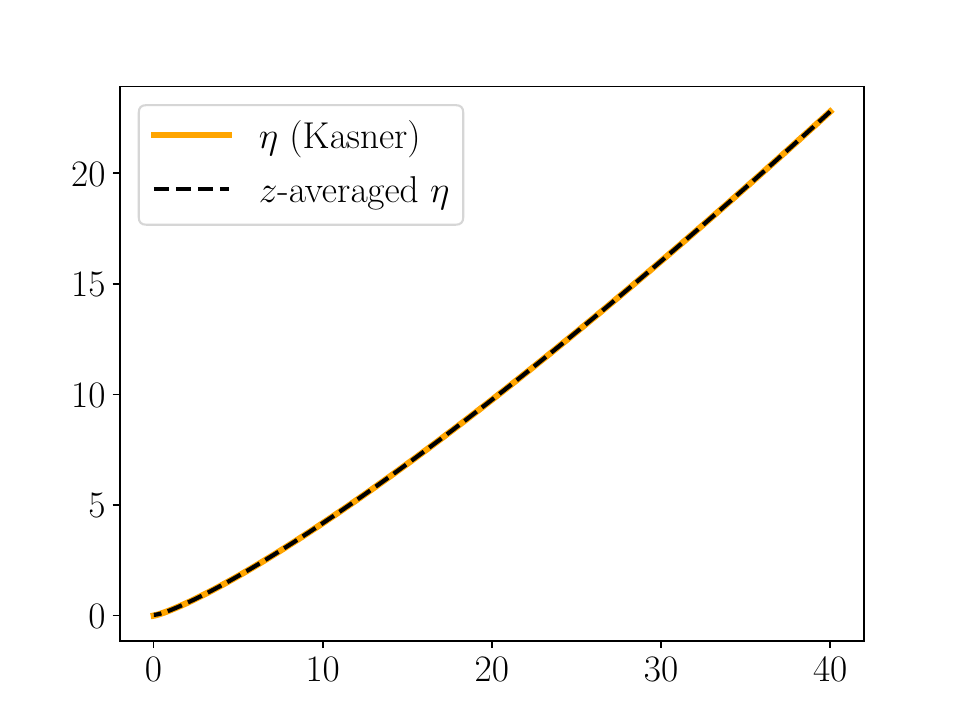}\\
\includegraphics[width=6cm]{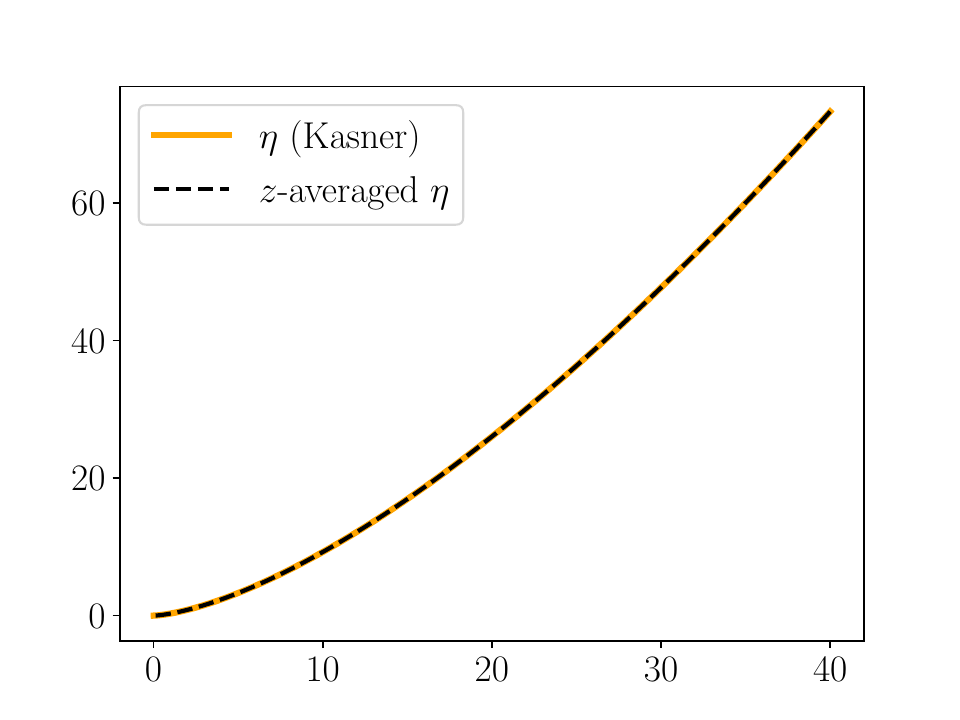}\includegraphics[width=6cm]{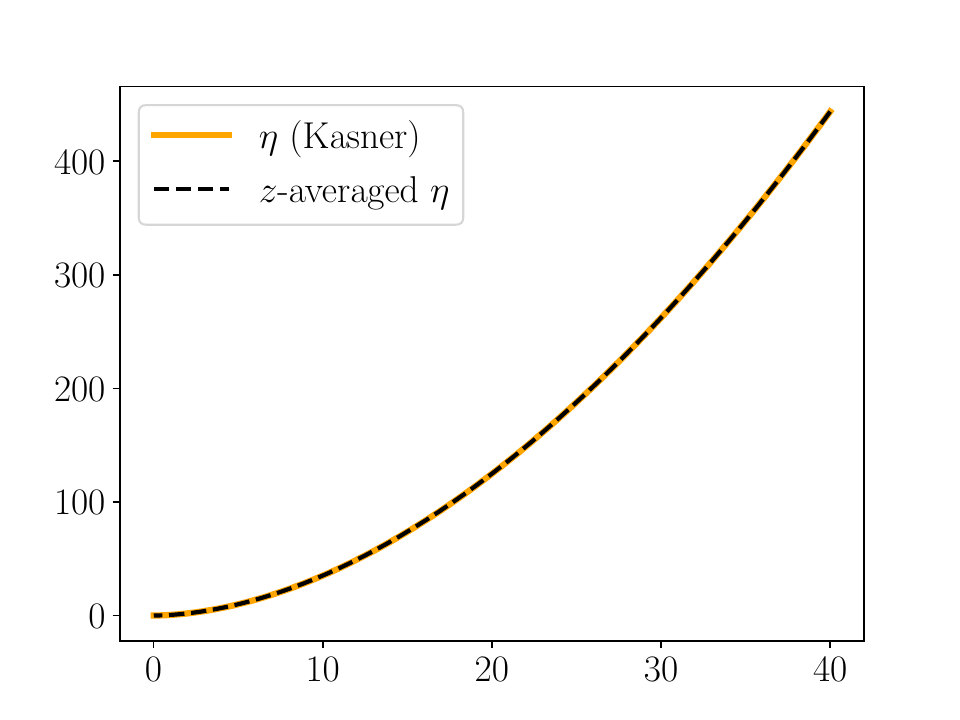}\\
\caption{Plots of $\eta$ of the best fitting kasner solution and $z$-averaged
$\eta$ of the solution (indicated as $\bar\eta$), for six solutions in \autoref{table_conv_series_J03}. From left to
right, from top to bottom, the solutions corresponding to $N_h = 22, 34, 46, 58,
70, 82.$}\label{figure_Kasner_fitting}
\end{figure}

For every solution in \autoref{table_conv_series_J03} we compute several
physical relevant quantities. The results are shown in Table \ref{table_quantities_series_J03}.
\begin{table}[hb!]
\captionsetup{margin=1cm}
\begin{center}
{\footnotesize
\begin{tabular}{ccccccrc}
\hline
$N_h$ & $L$ & $M$ & $\alpha$ (from $M$) & $\alpha$ (from $V$) & $|\Omega_{\cal H}|$ \\
\hline 
22 & 17.576096 & 2.022813 & 0.4604 & 0.4604 & 7.5128$\times 10^{-2}$ \\
24 & 16.111421 & 2.023039 & 0.5023 & 0.5023 & 7.5363$\times 10^{-2}$ \\
26 & 14.872081 & 2.023303 & 0.5442 & 0.5442 & 7.5632$\times 10^{-2}$ \\
28 & 13.809790 & 2.023609 & 0.5861 & 0.5861 & 7.5937$\times 10^{-2}$ \\
30 & 12.889137 & 2.023963 & 0.6281 & 0.6281 & 7.6280$\times 10^{-2}$ \\
32 & 12.083566 & 2.024372 & 0.6701 & 0.6701 & 7.6666$\times 10^{-2}$ \\
34 & 11.372768 & 2.024842 & 0.7122 & 0.7122 & 7.7098$\times 10^{-2}$ \\
36 & 10.740948 & 2.025380 & 0.7543 & 0.7543 & 7.7584$\times 10^{-2}$ \\
38 & 10.175635 & 2.025995 & 0.7964 & 0.7964 & 7.8133$\times 10^{-2}$ \\
40 & 9.666853  & 2.026696 & 0.8386 & 0.8386 & 7.8753$\times 10^{-2}$ \\
42 & 9.206527  & 2.027492 & 0.8809 & 0.8809 & 7.9452$\times 10^{-2}$ \\
44 & 8.788048  & 2.028397 & 0.9233 & 0.9233 & 8.0227$\times 10^{-2}$ \\
46 & 8.405959  & 2.029423 & 0.9657 & 0.9657 & 8.1073$\times 10^{-2}$ \\
48 & 8.055711  & 2.030586 & 1.0083 & 1.0083 & 8.1994$\times 10^{-2}$ \\
50 & 7.733482  & 2.031906 & 1.0510 & 1.0510 & 8.3042$\times 10^{-2}$ \\
52 & 7.436041  & 2.033406 & 1.0938 & 1.0938 & 8.4327$\times 10^{-2}$ \\
54 & 7.160632  & 2.035117 & 1.1368 & 1.1368 & 8.6007$\times 10^{-2}$ \\
56 & 6.904895  & 2.037073 & 1.1801 & 1.1801 & 8.8199$\times 10^{-2}$ \\
58 & 6.666795  & 2.039311 & 1.2236 & 1.2236 & 9.0851$\times 10^{-2}$ \\
60 & 6.444569  & 2.041876 & 1.2673 & 1.2673 & 9.3611$\times 10^{-2}$ \\
62 & 6.236679  & 2.044816 & 1.3115 & 1.3115 & 9.5820$\times 10^{-2}$ \\
64 & 6.041783  & 2.048199 & 1.3560 & 1.3560 & 9.6708$\times 10^{-2}$ \\
66 & 5.858699  & 2.052122 & 1.4011 & 1.4011 & 9.5882$\times 10^{-2}$ \\
68 & 5.686384  & 2.056737 & 1.4468 & 1.4468 & 9.4009$\times 10^{-2}$ \\
70 & 5.523916  & 2.062283 & 1.4933 & 1.4933 & 9.3474$\times 10^{-2}$ \\
72 & 5.370474  & 2.069114 & 1.5411 & 1.5411 & 9.8604$\times 10^{-2}$ \\
74 & 5.225326  & 2.077756 & 1.5905 & 1.5905 & 1.1500$\times 10^{-1}$ \\
76 & 5.087817  & 2.088970 & 1.6423 & 1.6423 & 1.4767$\times 10^{-1}$ \\
78 & 4.957360  & 2.103970 & 1.6977 & 1.6975 & 1.9812$\times 10^{-1}$ \\
80 & 4.833426  & 2.125070 & 1.7586 & 1.7585 & 2.6153$\times 10^{-1}$ \\
82 & 4.715538  & 2.158354 & 1.8308 & 1.8306 & 3.2765$\times 10^{-1}$ \\
84 & 4.603263  & 2.242520 & 1.9486 & 1.9482 & 4.2784$\times 10^{-1}$ \\
\hline
\end{tabular}
}
\caption{\label{table_quantities_series_J03} Komar mass ($M$), Kasner exponent
($\alpha$, computed with two different methods), and horizon angular velocity
($|\Omega_{\cal H}|$) obtained for the solutions in the series of \autoref{table_conv_series_J03}.}
\end{center}
\end{table}
The mass $M$ is computed as the integral $M(\rhomax)$. The horizon's angular
velocity is computed as the averaged value of $\Omega(\rho=0)$ in one of the
horizons, \footnote{The computation of $\Omega$ is singular at the horizon; we
compute $\Omega$ strictly in the interior and get the value of $\Omega(\rho=0)$
by simple linear extrapolation from the first and second internal gridpoints.}.
The Kasner exponent is computed in two different ways: from from the Mass, as
$4M/L$, and from the asymptotic behaviour of the function $V$; more precisely,
as the slope of a linear regression of $\ln(V)$ as function of $\ln(\rho)$ in
the {\em asymptotic region} of the computational domain, see \eqref{kasnersol}, defined arbitrarily as the third of the grid points closest to the outer boundary. The \autoref{table_conv_series_J03} shows that the two values obtained for the Kasner exponent are in very good agreement. To visualize the excellent fitting of the Kasner model, we plot in \autoref{figure_Kasner_fitting} the $z$-averaged values of $\eta$ of our numerical solutions together with the $\eta$ of the corresponding Kasner solution.

\begin{figure}[h!]
\captionsetup{margin=1cm}
\centering
\includegraphics[width=6cm]{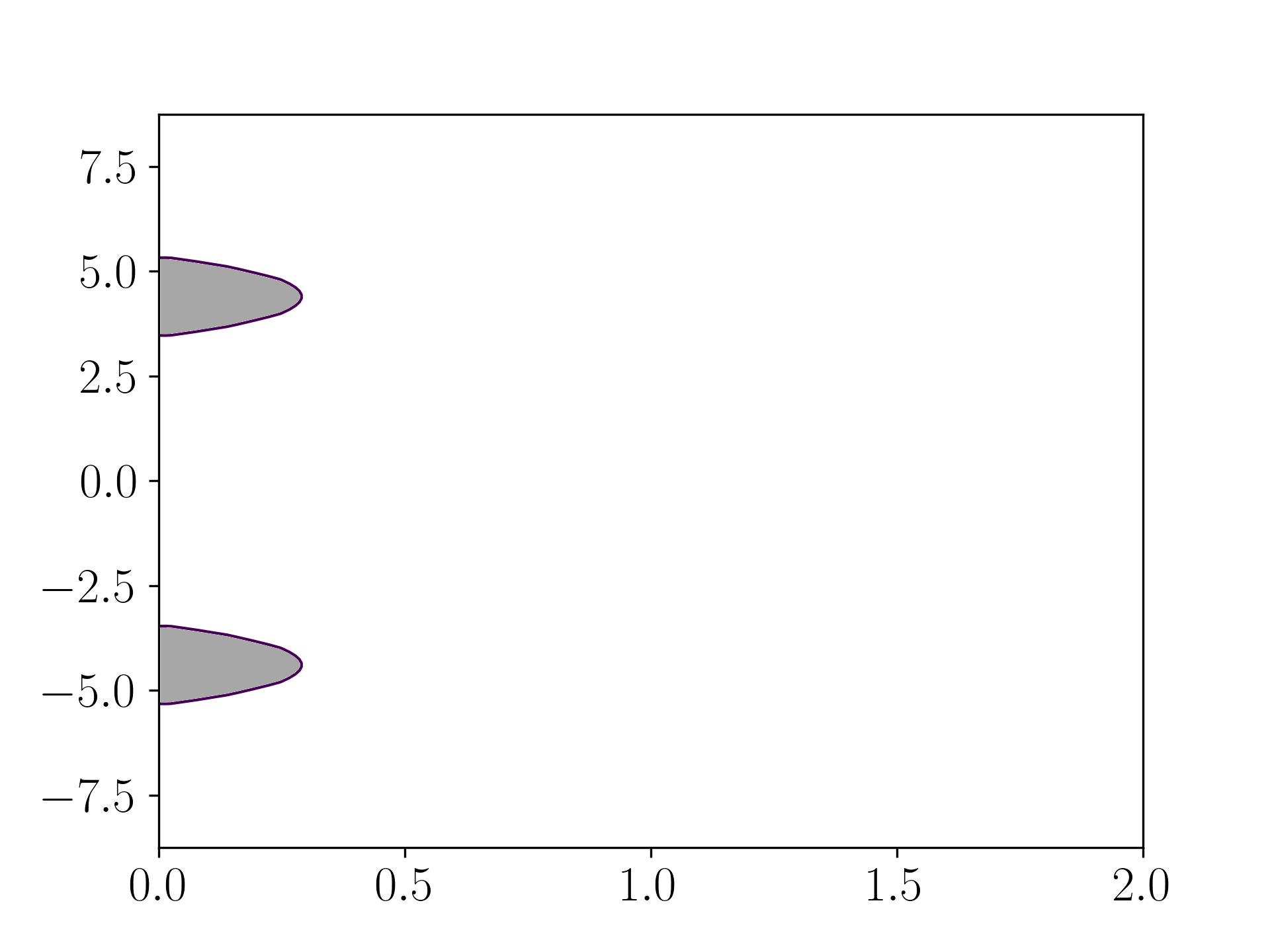}\includegraphics[width=6cm]{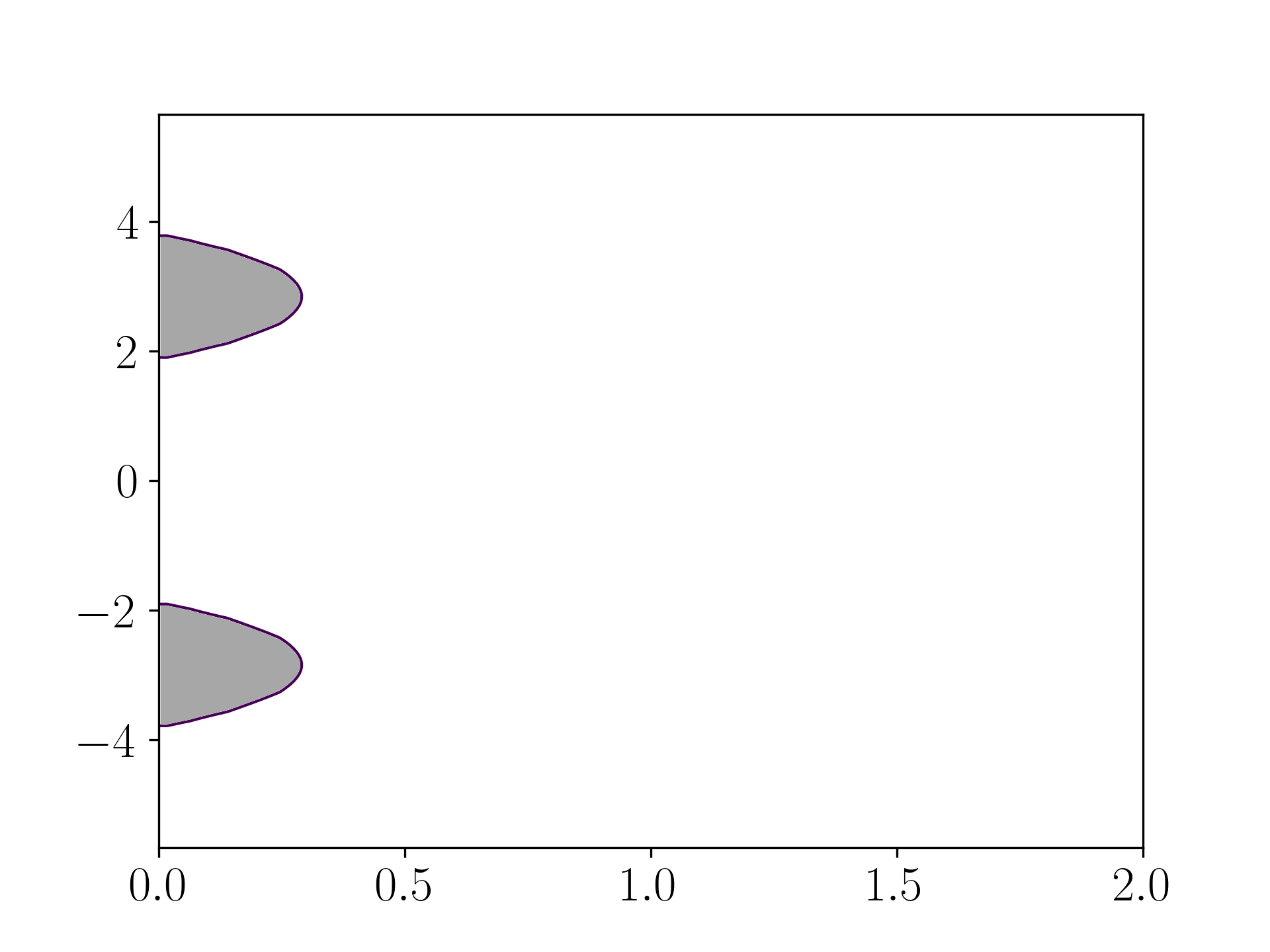}\\
\includegraphics[width=6cm]{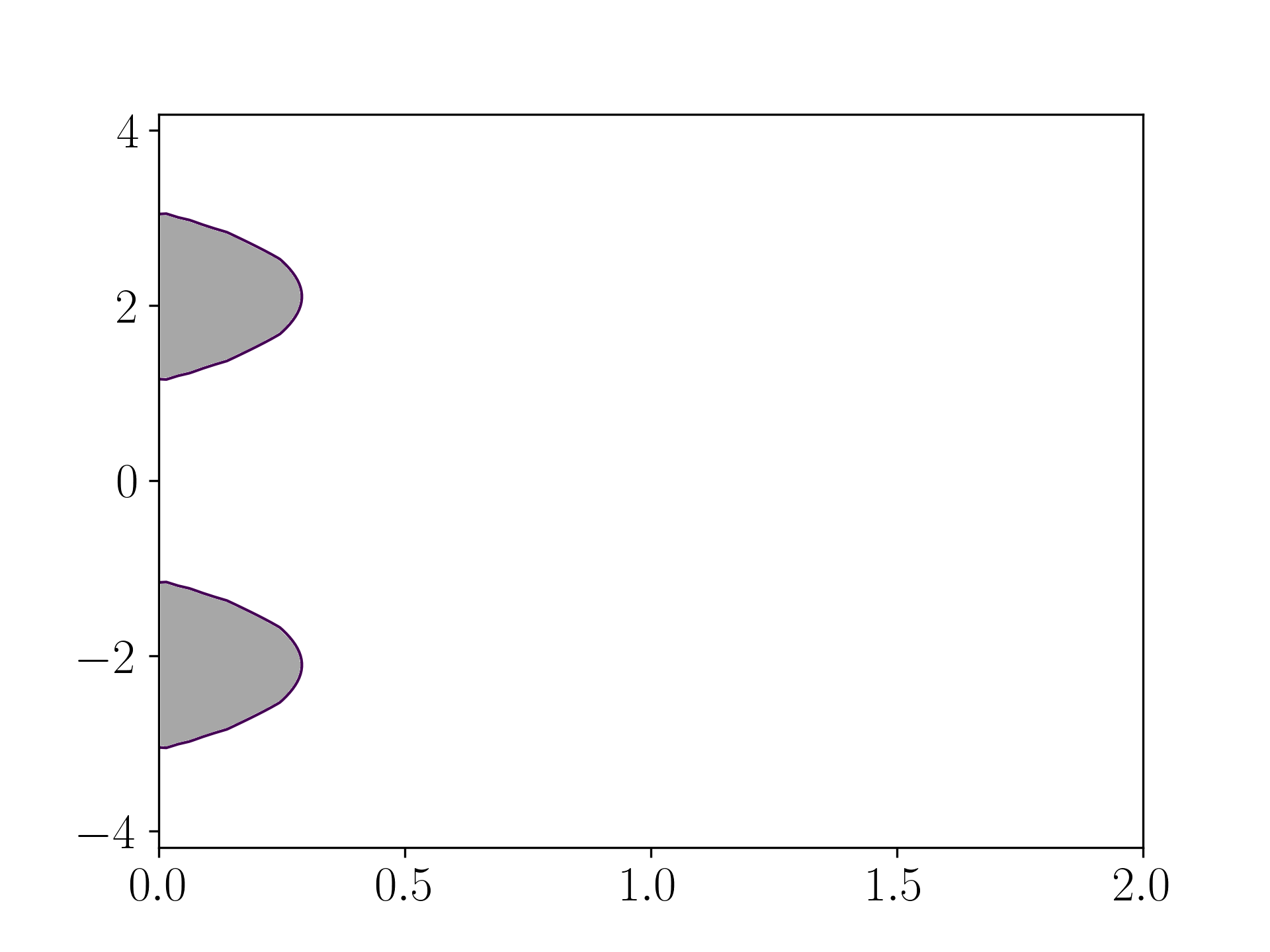}\includegraphics[width=6cm]{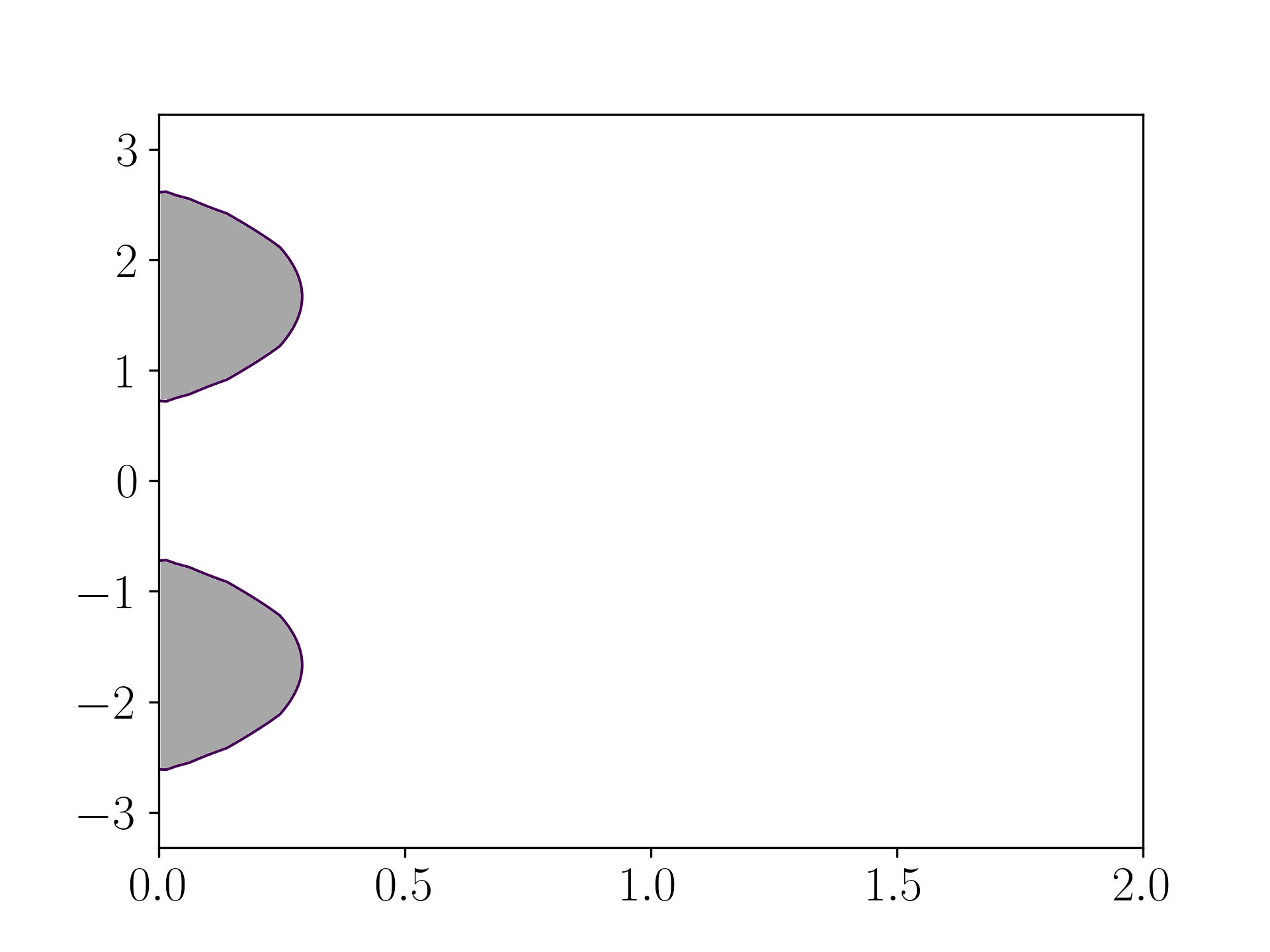}\\
\includegraphics[width=6cm]{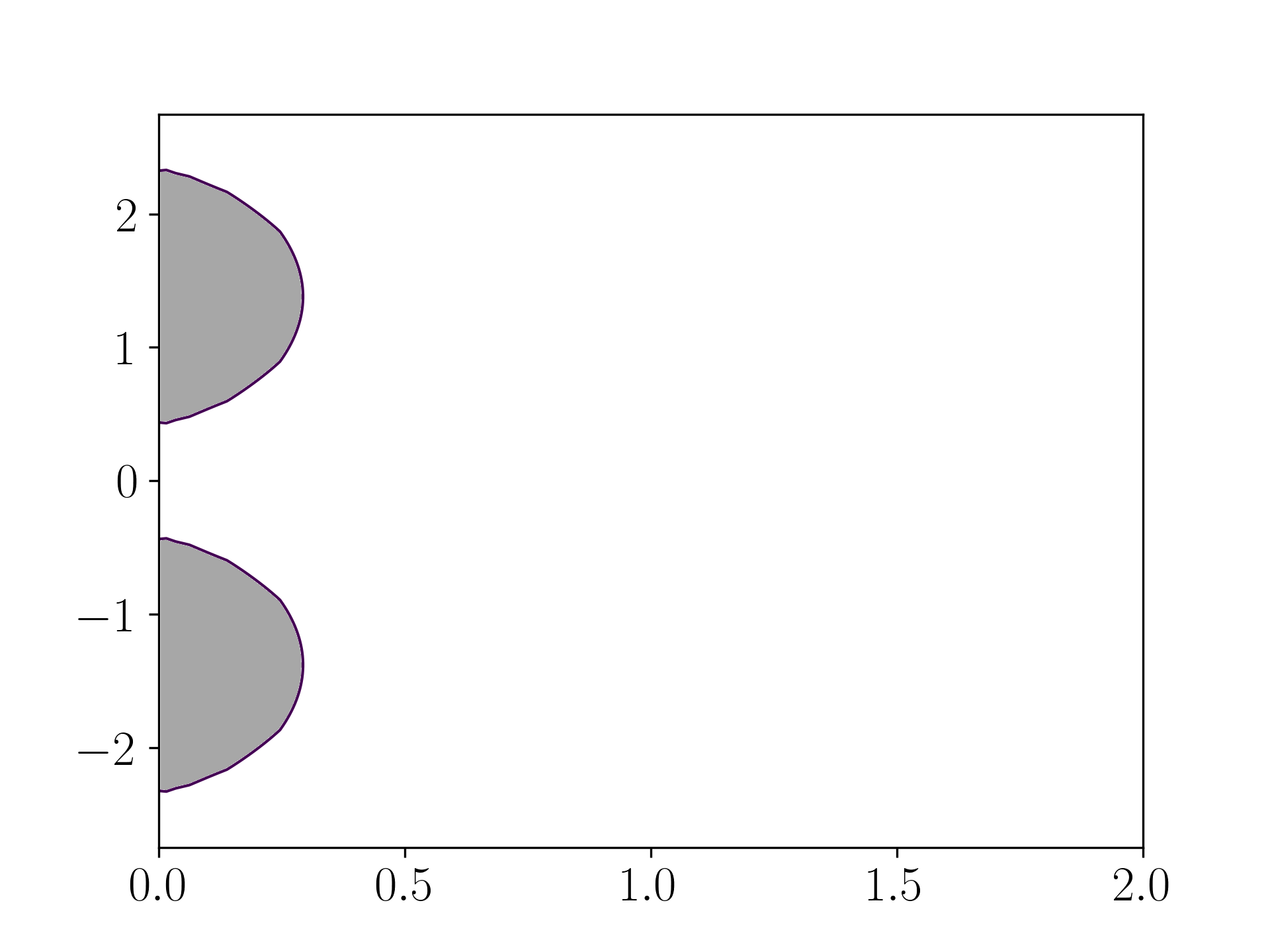}\includegraphics[width=6cm]{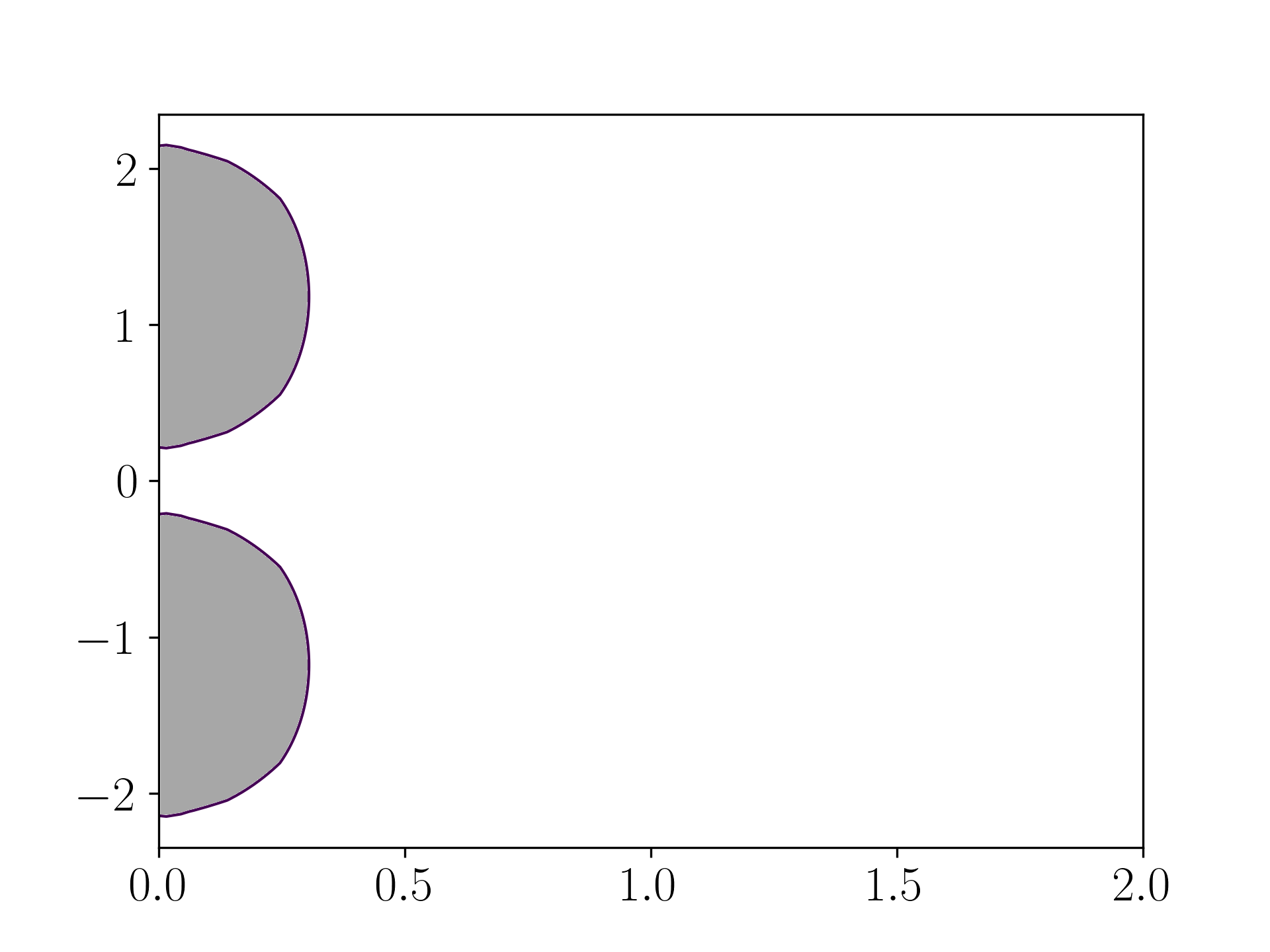}\\
\caption{Plots of a small region of the domain close to the symmetry axis showing the ergosphere for six solutions in \autoref{table_conv_series_J03}. From left to right, from top to bottom, the solutions corresponding to $N_h = 22, 34, 46, 58, 70, 82.$}\label{figure_ergospheres}
\end{figure}

We show plots of some of the ergospheres \footnote{By definition, an \emph{ergosphere} is the boundary of the region where $\partial_t$ is spacelike. In other words, the ergosphere is the boundary of the set $\{ V < 0 \}$, in terms of the function $V$ in \eqref{asm}. By continuity, $\partial_t$ is null on the ergosphere.} in \autoref{figure_ergospheres}, where we can see that in all cases they extend up to the same value of $\rho$. However they extend along the $z$-axis and covering a bigger region in $z$ of domain, and keeping always separated.





\section{Conclusions} \label{sec_conlusions}

In this paper we have presented a general scheme to construct numerical multi-horizon solutions in periodic set-ups. In particular, the introduction of the asymptotic boundary condition, \eqref{asymptotic_condition_sigma}, generalizes the previously proposed condition for single-horizon solutions in \cite{Peraza:2023boa}.

As a simple application, we give a detailed analysis of a binary configuration in a periodic setup, which we call counter-rotating solutions. They consist of two co-axial equidistant black holes with opposite spins but otherwise identical. These solutions have the advantage that the asymptotic region is that of a Kasner solution \eqref{kasnersol}.

We presented several properties, including the mass, the angular velocity of the horizons, and the ergospheres. Our study suggests that all the main features of the solutions presented here are also shown -- qualitatively -- for all values of $J$ that are not close to the extreme horizon case. Although in the present article we have shown in full detail runs for $J= 0.3$, we also performed runs with $J= 0.25$ and $J= 0.5$, obtaining the same qualitative results regarding convergence, asymptotic behaviour and angular velocity profile.

In comparison with the single horizon case, discussed in detail in \cite{Peraza:2023boa}, we show that the lower bound for $L=8m$ for the existence of solutions is not present in the binary case \footnote{Here we have the factor of $8$ instead of $4$ as in \cite{Peraza:2023boa}, since we have \emph{two} horizons per period.}. Indeed, in \autoref{table_quantities_series_J03} we construct solutions well below that threshold.

We see that a slight violation of the equidistant condition between the horizons leads to the breaking of the regularity due to the appearance of struts between them; see \autoref{table_conv_assymetric_sol}. 

Also, in view of \autoref{table_quantities_series_J03}, as the horizons get closer, the values of $\alpha$ tend monotonically to $2$. A Kasner solution with $\alpha = 2$ corresponds to the ``Boost'' solution, which is a quotient of the Rindler wedge, see \cite{Reiris2018a}. A still open conjecture presented on that work is the statement that a solution with the asymptotic behaviour of the Boost solution is the Boost solution. 

A natural question, in the context of the present work, is whether we can identify a physical mechanism by means of which the limit $\alpha = 2$ is not reachable from the counter-rotating configurations. Indeed, our results are consistent with the conjecture, in the following sense. From the numerical simulations, there appears to be a steep growth in the absolute value of the horizon angular velocity, $|\Omega_\Ho|$, while the mass $M$ monotonically increases as the horizons get closer, see \autoref{table_quantities_series_J03}. 

From the numerical analysis, we can understand this apparent inconsistency by inspecting the \emph{physical} distance, $D$, between the black holes, as the \emph{coordinate distance} $L/2 - 2m$ goes to zero. We obtain that certain derivatives of $\omega$ near the axis overflow as $m \rightarrow L/4$, and therefore there is an obstruction in the limit $D \rightarrow 0$. It would be interesting to study the extent and the reason for this behavior in more detail.

\section*{Acknowledgements}

We would like to thank Martín Reiris for presenting the problem, for key discussions, and comments on the results. The authors also thank Dmitry Korotkin for useful comments and suggestions. O.O. was partly supported by the ``Secretaría de Ciencia y Tecnología, Universidad Nacional de Córdoba, grant number 33620230100211CB''. J.P. was partially funded by CAP PhD Scholarships, during the beginning of this project, in the context of his PhD. Research at Perimeter Institute is supported in part by the Government of Canada through the Department of Innovation, Science and Economic Development Canada and by the Province of Ontario through the Ministry of Colleges and Universities. This work was supported by the Simons Collaboration on Celestial Holography.

\providecommand{\noopsort}[1]{}\providecommand{\singleletter}[1]{#1}%

\end{document}